\documentclass{article}
\usepackage{graphicx}
\usepackage{amssymb}
\usepackage{amsmath}
\usepackage{amscd}
\usepackage{mathrsfs}
\usepackage{latexsym}

\title{\textbf{Outline of a Generalization and a Reinterpretation of Quantum Mechanics Recovering Objectivity}}
\author{Claudio Garola,\footnote{Department of Mathematics and Physics, University of Salento, Italy; e-mail: garola@le.infn.it} %\quad Marco Persano,\footnote{Department of Mathematics and Physics, University of Salento, Italy; e-mail: persano@le.infn.it} 
\quad
Sandro Sozzo\footnote{School Of Management and Institute IQSCS, University of Leicester, Leicester, United Kingdom; e-mail: ss831@le.ac.uk}, \quad
Junde Wu\footnote{Department of Mathematics, Zhejiang University, P. R. China; e-mail: wjd@zju.edu.cn 
}}
\begin{document}  
\maketitle  
\begin{abstract}
The ESR model has been recently proposed in several papers to offer a possible solution of the problems raising from the nonobjectivity of physical properties in quantum mechanics (QM) (mainly the \emph{objectification problem} of the quantum theory of measurement). This solution is obtained by embodying the mathematical formalism of QM into a broader mathematical framework and reinterpreting quantum probabilities as conditional on detection rather than absolute. We provide a new and more general formulation of the ESR model and discuss time evolution according to it, pointing out in particular that both linear and nonlinear evolution may occur, depending on the physical environment.

\vspace{.2cm}
\end{abstract}

\section{Introduction\label{intro}}
It is well known that the standard interpretation of quantum mechanics (QM), though empirically successful, is a source of problems and paradoxes. One can avoid these difficulties by adopting a purely statistical interpretation of QM \cite{ba70}, but at the expense of accepting that QM has nothing to say about single items of physical systems (briefly, \emph{individual objects}). If one maintains instead that QM refers to individual objects and their properties,\footnote{This position is called ``realistic'' by some authors \cite{blm96}. It expresses, however, a very weak form of realism, which does not assume any \emph{a priori} model for individual objects and their properties and does not imply ontological commitments about the theoretical entities of QM (one could indeed interpret individual objects as activations of preparation procedures \cite{l83}). Such a weak form of realism is obviously implied by stronger realistic interpretations and/or modifications of QM, as Bohm's theory, many-worlds interpretation, GRW theory, etc.} as we will do in the following, then the \emph{objectification problem} arises which makes it difficult to work out a consistent quantum theory of measurement \cite{blm96}. The deep root of this problem is \emph{nonobjectivity} of physical properties,\footnote{For the sake of simplicity, we consider the notions of physical system, physical property and state as primitive in this section. We note, however, that physical properties can be intuitively interpreted as {\it dichotomic} observables, which can be measured obtaining one of two possible outcomes (often labeled {\it yes} and {\it no}).} which intuitively means that there are in QM physical properties that may be brought into existence by a measurement but do not preexist to it (\emph{ibid.}; see also \cite{me93}). 

Nonobjectivity is strongly supported by several theorems (often dubbed ``no--go'' theorem), as Bell--Kochen--Specker's \cite{b66,ks67}, which establishes that QM is a \emph{contextual} theory, and Bell's \cite{b64}, which establishes that QM is a \emph{nonlocal}, {\it i.e.}, contextual at a distance, theory.  Both these theorems imply nonobjectivity, for they state that the outcome of the measurement of a physical property $F$ on an individual object $\alpha$ may depend in QM not only on $F$ and on the state $S$ of $\alpha$, but also on the measurement context, even if the measurement that is performed is assumed to be \emph{exact} (efficiency 1, no flaws or random errors in the measuring apparatus). Further support to nonobjectivity is then provided by the results of experiments, as Aspect's \cite{aetal82,a82a} and similar successive experiments (see \cite{ge05} for a broad bibliography on this topic) which are usually interpreted as showing the nonlocality of QM.

Nonobjectivity, however, has many puzzling consequences besides the objectification problem. For instance, it entails that the usual \emph{epistemic} notion of probability cannot be maintained in the case of quantum probabilities, which are necessarily nonepistemic (or \emph{ontic}). This implies in particular that some ambiguities occur in the interpretation of mixed states (or \emph{mixtures}) in QM.\footnote{Indeed, all mixtures are represented by density operators in QM, and every such operator admits infinitely many decompositions in terms of pure states. If a density operator represents a {\it proper} mixture, there exists a decomposition whose coefficients are interpreted as epistemic probabilities. If a density operator represents instead an {\it improper mixture}, all coefficients of its decomposition are to be interpreted as nonepistemic probabilities \cite{desp76,tb05,gs12}.\label{improper}} Moreover, nonobjectivity implies, according to many scholars, that a nonclassical logic (\emph{quantum logic}) must be adopted in the (observational) language of QM, formalizing the properties of a notion of \emph{quantum truth} different from classical truth \cite{dcgg04}. Furthermore, nonobjectivity is counterintuitive, as witnessed by the long--standing debate about wave--particle duality. Indeed, it entails that no intuitive model for individual objects and their relationships can be constructed because such a model would imply that the physical properties of an individual object are independent of the measurement context.

Notwithstanding the problems outlined above, all early attempts at providing a hidden variables theory for QM, as Bohm's \cite{b52}, or operational foundations of QM, as the quantum logical or the C$^{*}$--algebra approach (see, \emph{e.g.}, \cite{blm96} for a short illustration of these approaches and related biography) preserved, more or less explicitly, the contextuality and the nonlocality of QM. Also some recent efforts of recovering the structure of QM from general principles and notions inspired by quantum information theory, as Zeilinger's foundational principle \cite{z99}, CBH theorem \cite{cbh03}, quantum Bayesianism \cite{cfs02a,cfs02b,fs04}, etc., either uphold a purely statistical interpretation or do not question contextuality and nonlocal correlations, which are instead considered as basic features and resources for quantum information processing. These approaches, of course, are supported by the theorems and experimental results mentioned above.

Philosophers of science, however, know that the interpretation of experimental data may be different in different theories. Moreover, one of us, together with various collaborators, has shown in some previous papers that the proofs of the ``no--go'' theorems rest on an implicit assumption on the range of validity of physical laws that is problematic in QM \cite{gp04,gp13}. If such an assumption is weakened, the proofs of the theorems cannot be completed. This suggests that ``objective'' interpretations of the formalism of QM cannot be \emph{a priori} excluded, even if they may imply some restrictions on the range of validity of QM. A \emph{semantic realism} (SR) interpretation of this kind was then proposed in several papers \cite{gp04,gs96a,gs96b,ga02,gshum} in which objectivity of physical properties was recovered at a semantic level, avoiding ontological commitments. More recently, two of us have proposed  an \emph{extended semantic realism} (ESR) model which modifies and generalizes the SR interpretation embodying the mathematical apparatus of QM into a broader mathematical setting that may admit an objective interpretation \cite{gs12,ga03,gs09,gs08,sg08,gs10,gs10b,gs10c,gps13,gserke,s13,g15}. 
There are two basic intuitive ideas underlying this model. The first of them is that, whenever a physical property $F$ of a physical system $\Omega$ is measured on an individual object $\alpha$ in a state $S$, the set of objective physical properties of $\alpha$ (which must be specified by the model but can be maintained to be non--void, in analogy with QM, see Sect. \ref{sub2}) may be such that $\alpha$ has nonzero probability of remaining undetected even if the measurement is {\it exact} (efficiency 1).
 This idea, which has some precedents in the literature that will be discussed in Sect. \ref{conclusions}, led us to associate a function depending on $F$ and $S$, called {\it detection probability}, with $\Omega$. Hence one must distinguish a {\it conditional on detection probability} of $F$ in $S$, which refers to the set of all individual objects that are detected by an exact measurement of $F$, and an {\it overall probability}, which refers to the set of all individual objects that are produced and is the product of detection and conditional on detection probabilities. The second idea underlying the ESR model is that the conditional on detection probability coincides with the probability of $F$ in $S$ provided by the quantum description of $\Omega$. Based on this idea, the mathematical apparatus of QM can be recovered within a broader mathematical framework in the ESR model. Nevertheless, the standard interpretation of the probability supplied by the Born rule as overall probability is not preserved, which has long ranging consequences. Indeed, the overall probability predicted by the ESR model is then different from the overall probability predicted by QM, except when the detection probability is identically 1 (but the difference between the predictions of the two theories can be very small and hidden by the lack of efficiency of real, non-exact measuring devices). Moreover, the reinterpretation of quantum probabilities provides the crucial tool for recovering an objective description of $\Omega$ whenever the detection probability satisfies suitable conditions, circumventing the ``no--go'' theorems. Indeed, all proofs of these theorems (be they based on inequalities \cite{b64,chsh69} or not \cite{me93,b66,ks67,ghsz82}) rest on the standard interpretation of quantum probability. 
 
 The ESR model, however, was presented in the papers quoted above mixing together a \emph{microscopic} (purely theoretical) and a \emph{macroscopic} (empirically interpreted) part. Moreover, time evolution was not discussed in this new theoretical framework. We therefore provide in the present paper a new presentation of the ESR model in which the macroscopic part is built up as a generalization and reinterpretation of QM (Sect. \ref{modello}). The microscopic part is then added as a noncontextual hidden variables theory intended both to provide a picture of the microworld that justifies the fundamental equation of the ESR model and to show by means of examples that the description of physical systems supplied by the ESR model may be noncontextual (Sect. \ref{hv}). We complete the ESR model by providing a general treatment of time evolution within the theoretical perspective introduced by it (Sect. \ref{evolution_meas}), and conclude this paper with a discussion of the advantages of the ESR model, its limits, and its relationships with the proposals of other scholars (Sect. \ref{conclusions}). Our presentation is preceded by a formal definition of the basic notion of objectivity and its connection with contextuality (Sect. \ref{objectivity}), and by a short summary of QM, intended to facilitate the comparison with the generalization and reinterpretation of QM introduced by the ESR model (Sect. \ref{standardqm}).

\section{Objectivity versus contextuality\label{objectivity}}
We have supplied an intuitive definition of objectivity in Sect. \ref{intro}. Since nonobjectivity of QM is the basic issue that motivated our research, let us make our definition more precise.

As in Sect. \ref{intro}, let us consider for the moment the notions of {\it physical property} and {\it state} as primitive (see footnote \ref{improper}; in QM and in the ESR model these notions will be operationally interpreted in terms of more primitive entities, see Sects. \ref{sub1} and \ref{esrsub1}, respectively). Let ${\mathscr T}$ be a physical theory describing the physical system $\Omega$, let $F$ be a physical property of $\Omega$ and let $\alpha$ be an individual object in a state $S$. A measurement of $F$ on $\alpha$, that we will suppose to be exact from now on, yields one of two possible outcomes (say {\it yes} and {\it no}, consistently with footnote 2). Such an outcome may be prefixed or not, according to ${\mathscr T}$, for the given measurement context. Then we say that $F$ is {\it objective} for $\alpha$ in ${\mathscr T}$ if and only if (iff) the following conditions hold.

\vspace{1mm}
\noindent
VD ({\it value definiteness}).  The outcome of a measurement of $F$ on $\alpha$ is prefixed in ${\mathscr T}$ for every measurement context.

\vspace{1mm}
\noindent
NC ({\it noncontextuality}). If the outcome of a measurement of $F$ on $\alpha$ is prefixed in ${\mathscr T}$ for some measurement contexts, then it is the same for each of these contexts.

\vspace{1mm}
\noindent
The definition above has some interesting consequences. Indeed, whenever the outcome of a  measurement of $F$ on $\alpha$ is prefixed in ${\mathscr T}$ for a given measurement context, one can reinterpret it as the truth value, in that context, of the sentence $F(\alpha)$ that attributes $F$ to $\alpha$ ({\it yes=true}, {\it no=false}). If conditions VD and NC are fulfilled, such truth value is assigned for every measurement context and does not depend on the context. Hence one can briefly say that, if $F$ is objective in ${\mathscr T}$, then a  measurement of $F$ reveals a  preexisting truth value of $F$.

Let us come to ${\mathscr T}$. We say that ${\mathscr T}$ is {\it contextual}  whenever some physical properties exist such that condition NC is not fulfilled. We say that ${\mathscr T}$ is {\it objective} if, for every state $S$ and individual object $\alpha$, every physical property is objective. It is evident that, if ${\mathscr T}$ is contextual, then it is nonobjective. Whenever ${\mathscr T}$ is nonobjective, instead, it may be noncontextual, for it could simply occur that there are physical properties in ${\mathscr T}$ such that condition VD is not fulfilled. Briefly, contextuality implies nonobjectivity, but the converse implication does  not hold.

\section{Recalls of standard QM\label{standardqm}}
We provide in this section a presentation of the basic notions of QM that will be used in the rest of this paper, with the aim of making the generalization and reinterpretation of QM in Sect. \ref{modello} as immediate and transparent as possible.

\subsection{Fundamental physical entities\label{sub1}}
A \emph{physical system} $\Omega$ can be theoretically described in QM by a triple $({\mathscr S},{\mathscr O},p)$, with $\mathscr S$ a set of \emph{states}, $\mathscr O$ a set of \emph{observables} and $p$ a \emph{probability mapping}.

The set $\mathscr S$ is partitioned into a set $\mathscr P$ of \emph{pure states} and a set ${\mathscr S}\setminus{\mathscr P}$ of \emph{mixed states}, or \emph{mixtures} and, according to some authors, ${\mathscr S}\setminus{\mathscr P}$ must be further partitioned into a set $\mathscr M$ of \emph{proper} mixtures and a set ${\mathscr N}$ of \emph{improper} mixtures \cite{desp76}.

Coming to observables, let us denote by ${\mathbb B}(\Re)$ the set of all Borel sets of the real line $\Re$, and for every set $\Gamma$ let ${\mathbb P}(\Gamma)$ denote the power set of $\Gamma$. Then, every observable $A \in {\mathscr O}$ is associated with a set $\Xi_{A} \in {\mathbb B}(\Re)$ of \emph{possible values} of $A$ and a set 
\begin{equation} \label{quantumproperties}
{\mathscr E}_{A}=\{ E=(A,\Sigma): \ \Sigma \in {\mathbb P}(\Xi_A) \cap  {\mathbb B}(\Re)\}
\end{equation} 
of \emph{quantum properties}. Hence the set 
\begin{equation} \label{allquantumproperties}
{\mathscr E}=\{ E=(A,\Sigma): \ A \in {\mathscr O},   \Sigma \in {\mathbb P}(\Xi_A) \cap  {\mathbb B}(\Re) \}=\cup_{A \in {\mathscr O}}{\mathscr E}_{A}
\end{equation}
is called the \emph{set of all quantum properties of $\Omega$}.

Finally, the mapping $p$ maps ${\mathscr S} \times {\mathscr E}$ into the interval $[0,1]$ of $\Re$, and is such that, for every $S \in {\mathscr S}$ and $A \in {\mathscr O}$, the mapping which maps $\Sigma \in {\mathbb P}({\Xi_A})\cap {\mathbb B}(\Re)$ into $p(S, (A,\Sigma))\in [0,1]$ is a probability measure on $\Xi_A$. Hence, for every $S \in {\mathscr S}$ and $E=(A,\Sigma)\in {\mathscr E}$, $p(S,E)$ is called the \emph{probability of $E$ in $S$}.\footnote{If one puts ${\mathscr N}=\emptyset$, the above scheme could refer to classical and statistical mechanics as well. Of course, for every $S \in {\mathscr P}$ and $E\in {\mathscr E}$, $p(S,E)\in \{ 0,1\}$ in classical mechanics. Furthermore, $p(S,E)$ admits an epistemic interpretation in these theories, at variance with QM (Sect. \ref{intro}).}

\subsection{Empirical interpretation\label{sub2}}
The theoretical entities introduced in Sect. \ref{sub1} can be empirically interpreted on macroscopic physical entities according to the following scheme \cite{l83,bc81}.\footnote{According to a known epistemological perspective (\emph{received viewpoint} \cite{b53,h65} assigning an empirical interpretation of the theoretical entities implies establishing \emph{correspondence rules} connecting the \emph{theoretical} language of a physical theory with its \emph{observational} language. We do not deepen this philosophical issue here, but stress that, generally, not all theoretical entities of a theory may have a direct empirical interpretation.}

A physical system $\Omega$ is associated with a triple $(\Pi, {\mathscr R}, \nu)$, with $\Pi$ a set of \emph{preparation procedures}, $\mathscr R$ a set of \emph{exact dichotomic registering devices}, whose outcomes are $+1$ (or \emph{yes}) and $-1$ (or \emph{no}), and $\nu$ a mapping of $\Pi \times {\mathscr R}$ into $[0,1]$. For every $(\pi, r) \in \Pi \times {\mathscr R}$, $\nu(\pi,r)$ is the large number limit of the frequency of the outcome $+1$ of $r$ whenever $r$ is used to perform a series of registrations, each occurring (immediately) after an activation of $\pi$.

The mapping $\nu$ induces two equivalence relations $\equiv$ and $\approx$ on $\Pi$ and $\mathscr R$, respectively, as follows.

Let $\pi_1,\pi_2 \in \Pi$. Then, $\pi_1 \equiv \pi_2$ iff for every $r \in \mathscr R$, $\nu(\pi_1,r)=\nu(\pi_2,r)$.

Let $r_1,r_2 \in {\mathscr R}$. Then, $r_1 \approx r_2$ iff for every $\pi \in \Pi$, $\nu(\pi,r_1)=\nu(\pi,r_2)$.

Every state $S \in {\mathscr S}$ is empirically interpreted on an equivalence class $[\pi]_{\equiv} \in \Pi/_{\equiv}$, and every quantum property $E=(A,\Sigma) \in {\mathscr E}$ is empirically interpreted on an equivalence class $[r]_{\approx} \in {\mathscr R}/_{\approx}$. Measuring $E$ in $S$ thus means applying a registering device in $[r]_{\approx}$ after activating a preparation procedure in $[\pi]_{\equiv}$, obtaining one of the outcomes {\it yes} and {\it no}.

Finally, the probability mapping $p$ is empirically interpreted on the mapping $\tilde{\nu}$ canonically induced by $\nu$ on $\Pi/_{\equiv} \times {\mathscr R}/_{\approx}$. More explicitly, for every $S \in {\mathscr S}$ and $E \in {\mathscr E}$ corresponding to $[\pi]_{\equiv} \in \Pi/_{\equiv}$ and $[r]_{\approx} \in {\mathscr R}/_{\approx}$, respectively, $p(S,E)\longrightarrow\tilde{\nu}([\pi]_{\equiv},[r]_{\approx})=\nu(\pi,r)$.

The above empirical interpretation is sufficient for our aims in this paper. It is easy to see, however, that it can be extended to observables. In this case quantum properties can be seen as special examples of observables: to be precise, dichotomic observables.\footnote{It is well known that the attempt at describing the dichotomic registering devices (or, more generally, the apparatuses corresponding to observables) in QM, together with their interaction with the physical system $\Omega$, raises the objectification problem mentioned in Sect. \ref{intro}. More specifically, nonobjectivity transfers to the macroscopic level, as illustrated by famous paradoxes. We avoid such problem here by adopting the above straightforward empirical interpretation of the theoretical entities of QM on the macroscopic entities in $\Pi$ and ${\mathscr R}$, as usual in elementary QM. Of course, in this presentation the question of whether QM can describe such entities and their interaction with $\Omega$ (that is, ultimately, the question of the universality of QM \cite{blm96}) remains unanswered. We come back on this issue in Sect. \ref{evolution_meas}.\label{notemeasurement}}

By considering explicitly individual objects, the empirical interpretation of QM can be further extended. Indeed, every activation of a preparation procedure $\pi$ can be assumed to prepare an individual object. Hence, when studying a physical system $\Omega$, one can introduce the set ${\mathscr U}$ of all individual objects (\emph{i.e.}, the set of all items of $\Omega$ that have been prepared) and, for every state $S$, the set $\textrm{ext}S \subset {\mathscr U}$ of all individual objects prepared by activating preparation procedures in the equivalence class corresponding to $S$. Then one says that an individual object $\alpha \in {\mathscr U}$ is in the state $S$ iff $\alpha \in \textrm{ext}S$. Moreover the family $\{\textrm{ext}S\}_{S \in {\mathscr S}}$ is a partition of $\mathscr U$. 

Let us consider now the objectivity issue. If the standard interpretation of QM is accepted, the ``no--go'' theorems mentioned in Sect. \ref{intro} show that QM is contextual in the sense specified in Sect. \ref{objectivity}. Hence QM is nonobjective (see again Sect. \ref{objectivity}).

 It is important to observe, however, that also objective quantum properties occur in QM. In fact a quantum property $E$ is objective, in the sense established in Sect. \ref{objectivity}, for every individual object $\alpha$ in the state $S$ (briefly, $E$ is objective in the state $S$) iff $p(S,E)$ is either 1 or 0 \cite{gp13}. This conclusion agrees with the definition of objectivity introduced in \cite{blm96}. If $p(S,E)=1$ ($0$) one then says that $E$ is \emph{possessed} (\emph{not possessed}) by $\alpha$. Whenever $0 \ne p(S,E) \ne 1$, instead, $E$ is nonobjective for every individual object $\alpha$ in the state $S$. Hence, if a measurement of $E$ on $\alpha$ yields the outcome \emph{yes} (\emph{no}), one can say that $\alpha$ \emph{displays} (\emph{does not display}) $E$ in the measurement, but the sentence $E(\alpha)$ has no truth value before the measurement. Every statement asserting that $\alpha$ possesses (does not possess) $E$ is meaningless in this case.\footnote{We recall that this position is weakened by the \emph{modal interpretations} of QM, which admit that, whenever $0\ne p(S,E)\ne 1$, $E$ could be objective for \emph{some} individual objects in the state $S$. Hence the modal interpretations of QM  distinguish between \emph{dynamical states} (that can be identified with the quantum states introduced above) and \emph{value states} (the value state of an individual object $\alpha$ representing, in our present terms, the set of all quantum properties that are objective for $\alpha$).} 

\subsection{Mathematical representation\label{sub3}}
We adopt in this paper the standard Hilbert space representation of the physical entities introduced in Sect. \ref{sub1}. Therefore the physical system $\Omega$ is associated with a complex separable Hilbert space $\mathscr H$ with scalar product $\langle \cdot | \cdot \rangle$. Then, in elementary QM each pure state $P\in {\mathscr P}$ is represented (up to a phase factor) by a vector $|\psi\rangle$ in the set ${\mathscr V}$ of all unit vectors of ${\mathscr H}$. More generally, states are represented by linear, positive, trace 1 operators (density operators) on $\mathscr H$. For the sake of simplicity we consider only the case in which no superselection rule occurs, so that the correspondence between the set of all states and the convex set ${\mathscr T}({\mathscr H})_{1}^{+}$ of all density operators in $\mathscr H$ is bijective. Pure states are then bijectively represented by the extremal elements of 
${\mathscr T}({\mathscr H})_{1}^{+}$ (hence the pure state $P$ is represented by the one--dimensional projection operator $\rho_{P}=|\psi\rangle\langle\psi|$), while no distinction occurs between the mathematical representations of proper and improper mixtures. Furthermore, every observable $A \in \mathscr O$ is represented by a self--adjoint operator $\widehat{A}$ whose spectrum is $\Xi_A$. Also the correspondence between the set of all observables and the set of all self--adjoint operators is supposed to be bijective. It follows that a quantum property $E=(A,\Sigma)$ is represented by the (orthogonal) projection operator $P^{\widehat{A}}(\Sigma)$ (equivalently, $P^{\widehat{A}}(X)$, with $X$ any Borel set of $\Re$ such that $X \cap \Xi_A=\Sigma$), with $P^{\widehat{A}}$ the spectral projection--valued (PV) measure on $\Re$ associated with $\widehat{A}$. Finally, for every state $S$ and quantum property $E=(A,\Sigma)$, the probability $p(S,E)$ is supplied by the Born rule
\begin{equation}
p(S,E)=\textrm{Tr}[\rho_S P^{\widehat{A}}(\Sigma)] ,
\end{equation}
where $\textrm{Tr}$ is the trace operation and $\rho_S$ the density operator representing $S$.

Whenever measurements are considered, it is usual in QM to assume that a subset exists of exact registering devices that perform measurements satisfying the L\"{u}ders rule (\emph{first kind}, \emph{ideal} measurements). To be precise, let $\alpha$ be an individual object in the state $S$. Then, an ideal first kind measurement of a quantum property $E=(A,\Sigma)$ on $\alpha$ which yields the outcome \emph{yes} transforms $S$ into a final state $S_{F}$ represented by the density operator
\begin{equation} \label{SN}
\rho_{S_{F}}=\frac{P^{\widehat{A}}(\Sigma)\rho_{S}P^{\widehat{A}}(\Sigma)}{\textrm{Tr}[P^{\widehat{A}}(\Sigma)\rho_{S}P^{\widehat{A}}(\Sigma)]} .
\end{equation}
The rule expressed by Eq. (\ref{SN}) is often (somewhat improperly) referred in the literature as \emph{L\"{u}ders' postulate} \cite{bc81}.

\section{The ESR model\label{modello}}
As we have anticipated in Sect. \ref{intro}, the ESR model aims to provide a generalization and reinterpretation of QM which avoids nonobjectivity. The basic intuitive ideas underlying this model have been described in Sect. \ref{intro}.  We show in the next sections that these ideas, though very simple, lead to a deep reinterpretation and enlargement of the formalism of QM.

\subsection{Fundamental physical entities\label{esrsub1}}
A physical system $\Omega$ is theoretically described in the ESR model by a quintuple $({\mathscr S}, {\mathscr O}_{0},p^t,p^d,p)$, with $\mathscr S$ a set of \emph{states}, ${\mathscr O}_0$ a set of \emph{generalized observables} and $p^t$, $p^d$, $p$ \emph{probability mappings}.

{\it States}. The set $\mathscr S$ corresponds to the set denoted by the same symbol in QM (Sect. \ref{sub1}). Hence it is partitioned into a set $\mathscr P$ of \emph{pure states}, a set $\mathscr M$ of \emph{proper mixtures} and a set  $\mathscr N$ of \emph{improper mixtures}.

{\it Generalized observables}. Let us adopt the same conventions on symbols established in Sect. \ref{sub1}. Then, every generalized observable $A_0 \in {\mathscr O}_{0}$ corresponds to an observable $A \in {\mathscr O}$ of QM, and it is obtained from $A$ by adding a \emph{no--registration outcome} $a_0$ to the set $\Xi_A \in {\mathbb B}(\Re)$ of all possible values of $A$. Hence $\Xi_{A_0}=\Xi_{A} \cup \{ a_0 \}$ is the Borel set of all possible values of $A_0$ (we assume in the following that $a_0 \in \Re$, which is not restrictive: indeed, if $\Xi_A=\Re$, one can choose a bijective Borel function $f:\Re\longrightarrow \Xi_{f(A)}$ such that $\Xi_{f(A)}\subset \Re$, and replace $A$ by $f(A)$). By analogy with QM, every $A_0 \in {\mathscr O}_{0}$ is associated with a set 
\begin{equation}
{\mathscr F}_{A_0}=\{ F=(A_0, \Sigma): \ \Sigma \in {\mathbb P}(\Xi_{A_0}) \cap {\mathbb B}(\Re) \}
\end{equation}
of \emph{physical properties}, and the set 
\begin{equation}
{\mathscr F}_{0}=\{ F=(A_0, \Sigma): \ A_{0} \in {\mathscr O}_{0}, \Sigma \in {\mathbb P}(\Xi_{A_0}) \cap {\mathbb B}(\Re) \} =\cup_{A_0 \in {\mathscr O}_{0}} {\mathscr F}_{A_0}
\end{equation}
is called the \emph{set of all physical properties of $\Omega$}. In addition, we introduce the subset
\begin{equation} \label{ESRallphysicalproperties}
{\mathscr F}=\{ F=(A_0, \Sigma): \ A_0 \in {\mathscr O}_{0} \ \Sigma \in {\mathbb P}(\Xi_{A_0}\setminus \{ a_0 \}) \cap {\mathbb B}(\Re) \} \subset {\mathscr F}_{0}.
\end{equation}
 Equation (\ref{ESRallphysicalproperties}) implies that a bijective mapping 
 \begin{equation}\label{bijectivemapping}
 g: (A_0,\Sigma)\in {\mathscr F} \longrightarrow (A,\Sigma) \in {\mathscr E}
\end{equation} 
  exists which maps $\mathscr F$ into the set $\mathscr E$ of all quantum properties of $\Omega$ (Sect. \ref{sub1}).

{\it Probability mappings}. The mapping $p^{t}$ maps ${\mathscr S}\times{\mathscr F}_{0}$ into the interval $[0,1]\subset \Re$. The mappings $p^{d}$ and $p$ map instead ${\mathscr S}\times{\mathscr F}$ into $[0,1]$.  

For every $S \in {\mathscr S}$ and $A_0 \in {\mathscr O}_{0}$, the mapping
\begin{equation}\label{pt}
\widehat{{p}^t}: \Sigma \in {\mathbb P}(\Xi_{A_0}) \cap {\mathbb B}(\Re) \longrightarrow p^t(S,(A_0,\Sigma))\in [0,1]
\end{equation}
is a probability measure on $\Xi_{A_0}$. Hence, for every $S \in \mathscr S$ and $F=(A_0,\Sigma)\in {\mathscr F}_{0}$, $p^{t}(S,F)$ is called \emph{the overall probability of $F$ in $S$}. 

For every $S \in \mathscr S$ and $A_0 \in {\mathscr O}_{0}$, the mapping
\begin{equation} \label{QM}
\widehat{p}: \Sigma \in {\mathbb P}(\Xi_{A_0}\setminus \{ a_0 \}) \cap {\mathbb B}(\Re)\longrightarrow p(S,(A_0,\Sigma))\in [0,1]
\end{equation}
is a probability measure on $\Xi_{A_0} \setminus \{ a_0 \}=\Xi_A$. Hence, for every $S \in \mathscr S$ and $F=(A_0,\Sigma)\in \mathscr F\subset {\mathscr F}_{0}$, $p(S,F)$ is called \emph{the conditional on detection probability of $F$ in $S$}.

For every $S \in {\mathscr S}$ and $F=(A_0,\Sigma)\in \mathscr F\subset {\mathscr F}_{0}$, the mapping $p^{d}$ is such that the following equation holds
\begin{equation} \label{formuladipartenza}
p^{t}(S,(A_0,\Sigma))= p^{d}(S,(A_0,\Sigma))p(S,(A_0, \Sigma)) \ .
\end{equation}
Hence, $p^{d}(S,F)$ is called \emph{the detection probability} of $F$ in $S$.

All these nouns are justified by the empirical interpretation to be discussed in the next section.

Finally, for every $F=(A_0, \Sigma)\in {\mathscr F}_{0}$, let us consider the {\it complementary property} $F^{c}=(A_0, \Xi_{A_0} \setminus \Sigma)$ of $F$. Then, the definition of $p^{t}$ implies that the following equation holds.
\begin{equation} \label{formuladipartenzaT}
p^{t}(S,(A_0,\Sigma))= 1-p^{t}(S,(A_0, \Xi_{A_0} \setminus \Sigma)) \ .
\end{equation}

\subsection{Empirical interpretation\label{esrsub2}}
The theoretical entities introduced in Sect. \ref{esrsub1} are empirically interpreted on macroscopic physical entities according to the following scheme.

The physical system $\Omega$ is associated with a quintuple $(\Pi, {\mathscr R}_{0},\nu^t,\nu^d,\nu)$. In this quintuple $\Pi$ is the same set of preparation procedures that occurs in the empirical interpretation of QM (Sect. \ref{sub2}). The set ${\mathscr R}_{0}$ is instead a set of exact (efficiency 1) registering devices with three possible outcomes, that we label $+1$, $0$ and $-1$, meaning that $0$ is the initial position of a pointer. Then, $\nu^t$, $\nu^d$ and $\nu$ are frequency functions which map $\Pi \times {\mathscr R}_{0}$ into $[0,1]$. For every $(\pi, {r}_{0}) \in \Pi \times {\mathscr R}_{0}$, $\nu^{t}(\pi,r_0)$ is the large number limit of the frequency of the outcome $+1$ of $r_0$ whenever $r_0$ is used to perform a series of registrations, each occurring after an activation of $\pi$; $\nu^{d}(\pi,r_0)$ is the complement to 1 of the large number limit of the frequency of the outcome $0$ of $r_0$ in the same series of registrations; $\nu(\pi,r_0)$ is the large number limit of the frequency of the outcome $+1$ of $r_0$ whenever only registrations of the series in which the outcome $0$ did not occur are considered.

The above definitions imply that the following equation holds
\begin{equation} \label{just1}
\nu^{t}(\pi,r_0)=\nu^{d}(\pi,r_0)\nu(\pi,r_0).
\end{equation}
The mappings $\nu^t$ and $\nu^d$ induce two equivalence relations $\equiv_{0}$ and $\approx_{0}$ on $\Pi$ and ${\mathscr R}_{0}$, respectively, as follows.

Let $\pi_1,\pi_2 \in \Pi$. Then, $\pi_1 \equiv_{0} \pi_2$ iff for every $r_0 \in {\mathscr R}_{0}$, $\nu^{t}(\pi_1,r_0)=\nu^{t}(\pi_2,r_0)$ and $\nu^{d}(\pi_1,r_0)=\nu^{d}(\pi_2,r_0)$.

Let $r_{01},r_{02} \in {\mathscr R}_{0}$. Then, $r_{01} \approx_{0} r_{02}$ iff for every $\pi \in \Pi$, $\nu^{t}(\pi,r_{01})=\nu^{t}(\pi,r_{02})$ and $\nu^{d}(\pi,r_{01})=\nu^{d}(\pi,r_{02})$.

Every state $S \in \mathscr S$ is then empirically interpreted on an equivalence class $[\pi]_{\equiv_{0}} \in \Pi/_{\equiv_{0}}$, and every physical property $F=(A_{0},\Sigma) \in {\mathscr F}\subset {\mathscr F}_{0}$ on an equivalence class $[r_{0}]_{\approx_{0}} \in {\mathscr R}_{0}/_{\approx_{0}}$. Measuring $F$ in $S$ then means applying a registering device in $[r_{0}]_{\approx_{0}}$ after activating a preparation procedure in $[\pi]_{\equiv_{0}}$. If one obtains the outcome $+1$, one says that the result is \emph{yes}; if one obtains the outcome $0$ or $-1$, one says that the result is \emph{no}.

Finally, the probabilities $p^t$, $p^d$ and $p$ are empirically interpreted on the mappings $\tilde{\nu^{t}}$,  $\tilde{\nu^{d}}$ and $\tilde{\nu}$ canonically induced on $\Pi/_{\equiv_{0}} \times {\mathscr R}_{0}/_{\approx_{0}}$ by $\nu^t$, $\nu^d$ and $\nu$, respectively. More explicitly, for every $S \in {\mathscr S}$ and $F \in {\mathscr F}\subset {\mathscr F}_{0}$ corresponding to $[\pi]_{\equiv_{0}} \in \Pi/_{\equiv_{0}}$ and $[r_{0}]_{\approx_{0}} \in {\mathscr R}_{0}/_{\approx_{0}}$, respectively, 
\begin{eqnarray}
p^{t}(S,F)\longrightarrow\tilde{\nu^{t}}([\pi]_{\equiv_{0}},[r_{0}]_{\approx_{0}})&=&\nu^{t}(\pi,r_{0}), \\
p^{d}(S,F)\longrightarrow\tilde{\nu^{d}}([\pi]_{\equiv_{0}},[r_{0}]_{\approx_{0}})&=&\nu^{d}(\pi,r_{0}), \\
p(S,F)\longrightarrow\tilde{\nu}([\pi]_{\equiv_{0}},[r_{0}]_{\approx_{0}})&=&\nu(\pi,r_{0}).
\end{eqnarray}

We must still supply an empirical interpretation of the properties in ${\mathscr F}_{0}\setminus{\mathscr F}$. To this end, let us observe that, if $F=(A_{0},\Sigma) \in {\mathscr F}_{0}\setminus{\mathscr F}$, then the complementary property $F^{c}$ belongs to ${\mathscr F}$. Hence $F$ is interpreted on the class $[r_{0}^{c}]_{\approx_{0}}$ of dichotomic registering devices corresponding to $F^{c}$. Measuring $F$ in $S$ thus means applying a registering device in $[r_{0}^{c}]_{\approx_{0}}$ after activating a preparation procedure in $[\pi]_{\equiv_{0}}$. If one obtains the outcome $+1$, one says that the result is \emph{no}; if one obtains the outcome $0$ or $-1$, one says that the result is \emph{yes}. It follows that the large number limit of the frequency of the outcome \emph{yes} in this kind of measurement is given by $1-\nu^{t}(\pi,r_0^{c})$. Therefore the probability $p^t$ is empirically interpreted as follows:
\begin{equation} \label{just2}
p^{t}(S,F)\longrightarrow 1-\tilde{\nu^{t}}([\pi]_{\equiv_{0}},[r_{0}^{c}]_{\approx_{0}})=1-\nu^{t}(\pi,r_{0}^{c}).
\end{equation}

Let us recall that the probabilities $p^d$ and $p$ are not defined on ${\mathscr S}\times ({\mathscr F}_{0}\setminus {\mathscr F})$. Hence, the empirical interpretation of states, physical properties and probabilities is now complete, which is sufficient for our aims in this paper. It could obviously be extended to observables, but we do not afford this task here for the sake of brevity. We observe instead that, at variance with QM, the measurement of a physical property in the ESR model is not a special case of the measurement of a generalized observable if \emph{yes} and \emph{no} are considered as its possible results. Indeed, the no--registration outcome does not occur explicitly as a separate outcome in this case.

By considering explicitly individual objects, the empirical interpretation provided above can be extended, as in QM. One can introduce the set $\mathscr U$ of all individual objects, the set $\textrm{ext}S\subset \mathscr U$, and the partition $\{ \textrm{ext}S \}_{S \in \mathscr S}$ as in Sect. \ref{sub2}. We instead cannot supply a criterion of objectivity in the ESR model at this stage, as we did in the case of QM, and postpone the discussion of objectivity in the ESR model to Sect. \ref{submicro2}.

\subsection{Basic assumptions\label{esrsub3}}
Equations (\ref{formuladipartenza}) and (\ref{formuladipartenzaT}) can now be considered as assumptions that are physically justified by the empirical interpretation in Sect. \ref{esrsub2}. To make them more transparent, we rewrite them as follows.

\vspace{1mm}
\noindent
\textbf{AX 1.} For every $S \in \mathscr S$, $F \in \mathscr F$,
\begin{equation}\label{ax1}
p^{t}(S,F)=p^{d}(S,F)p(S,F).
\end{equation}

\vspace{1mm}
\noindent
\emph{Physical justification}. Equation (\ref{just1}).

\vspace{1mm}
\noindent
\textbf{AX 2.} For every $S \in \mathscr S$, $F \in {\mathscr F}_{0}\setminus {\mathscr F}$,
\begin{equation}\label{f_bar_c}
p^{t}(S,F)=1-p^{t}(S,F^{c}).
\end{equation}

\vspace{1mm}
\noindent
\emph{Physical justification}. Equation (\ref{just2}).

\vspace{1mm}
\noindent
Because of AX 2 we will mainly consider physical properties in ${\mathscr F} \subset {\mathscr F}_{0}$ in the following.

The following statement is now introduced as a new {\it fundamental assumption} of the ESR model.

\vspace{1mm}
\noindent
\textbf{AX 3.} Let $P \in {\mathscr P}$ and $F \in {\mathscr F}$. Then the probability $p(P,F)$ coincides with the quantum probability $p(P,E)$, with $E$ the quantum property corresponding to $F$ via the mapping $g$ defined by Eq. (\ref{bijectivemapping}).

\vspace{1mm}
\noindent
\emph{Physical justification}. AX 3 implies that the ESR model embodies the basic mathematical formalism of QM. Hence this model does not formally conflict with QM, which is a fundamental requirement if one wants to take into account the outstanding empirical success of QM.

\vspace{1mm}
\noindent
AX 3 deeply modifies the interpretation of the mathematical formalism of QM. Indeed, consider the set $\textrm{ext}P$ of all individual objects in the pure state $P$. According to QM, whenever an exact measurement of a physical property $E$ is performed on an individual object $\alpha \in \textrm{ext}P$, detection always occurs and the quantum rules yield the overall probability that the outcome \emph{yes} is obtained. According to the ESR model, instead, whenever a measurement of $F=g^{-1}(E)$ is performed on $\alpha$, only the individual objects in a subset $(\textrm{ext}P)^{d} \subset \textrm{ext}P$ are detected, and the quantum rules yield the conditional on detection probability that the \emph{yes} result is obtained in a measurement whenever $\alpha \in (\textrm{ext}P)^{d}$ (Sect. \ref{esrsub1}).

It remains to stress that the detection probability $p^{d}(S,F)$ cannot be evaluated by using quantum rules. We have as yet no theory which enables us to predict it: hence it must be considered a parameter whose values are to be determined empirically case by case. We have proved elsewhere that some restrictions exist on its possible values if the ESR model is supposed to be objective \cite{gs08,gs10,gs10b}, and come back to this issue in Sects. \ref{submicro2} and \ref{conclusions}.

\subsection{Mathematical representation\label{esrsub4}}
The reinterpretation of quantum probabilities introduced by AX 3 has some important consequences. In particular, it entails that the mathematical formalism of QM must be extended if one wants to calculate overall probabilities. By introducing $p^{d}(S,F)$ into such formalism one can obtain the mathematical representations of states, generalized observables and physical properties that must be used in the ESR model to evaluate overall and conditional on detection probabilities, as follows.

\vspace{1mm}
\noindent
(i) \emph{The conditional on detection probability} (\emph{pure states only}). Let $F=(A_0, \Sigma) \in {\mathscr F}$ (hence $a_0 \notin \Sigma$) and $P \in \mathscr P$. Then AX 3 implies that, as far as $p(P,F)$ is concerned, $P$ can be represented as in QM. More explicitly, the physical system $\Omega$ is associated with a complex separable Hilbert space $\mathscr H$, $P$ is represented by a unit vector $|\psi\rangle\in {\mathscr V}\subset {\mathscr H}$ or by the one--dimensional projection operator $\rho_{P}=|\psi\rangle\langle\psi|$, and the latter representation is bijective if no superselection rule occurs (Sect. \ref{sub3}). Moreover, $A_0$ can be represented by the self--adjoint operator $\widehat{A}$ that represents, in QM, the observable $A \in {\mathscr O}$ from which $A_0$ is obtained (Sect. \ref{esrsub1}). Hence $F$ can be represented by the (orthogonal) projection operator $P^{\widehat{A}}(\Sigma)$ (equivalently, $P^{\widehat{A}}(X)$ with $X$ any Borel set of $\Re$ such that $X \cap \Xi_{A_{0}}=\Sigma$), where $P^{\widehat{A}}$ is the PV measure on $\Xi_A$ associated with $\widehat{A}$. Finally, the conditional on detection probability $p(P,F)$ can be calculated by using the standard quantum rule
\begin{equation} \label{prob_X_QM}
p(P,F)=Tr [\rho_{P} P^{\widehat{A}}(\Sigma)]=Tr[\rho_P \int_{\Sigma}P^{\widehat{A}(d \lambda)}].
\end{equation}
It follows in particular from Eq. (\ref{prob_X_QM}) that the mapping $\widehat{p}$ defined by Eq. (\ref{QM}) is a probability measure on $\Xi_{A_0}\setminus \{ a_0\}=\Xi_{A}$, as required in Sect. \ref{esrsub1}.

\vspace{1mm}
\noindent
(ii) \emph{The overall probability} (\emph{pure states only}). 
Bearing in mind the mathematical representations above and Eq. (\ref{ax1}), we obtain that, for every $P \in \mathscr P$ and $F=(A_0, \Sigma) \in {\mathcal F}$,
\begin{equation}\label{prob_X_W}
p^{t}(P,F)=p^{d}(P,F)Tr[\rho_{P} \int_{\Sigma} P^{\widehat{A}}(\mathrm{d}\lambda)]=Tr[\rho_{P}T_{P,A_{0}}(\Sigma)],
\end{equation}
with
\begin{equation}\label{newoperator}
T_{P,A_{0}}(\Sigma)=p^{d}(P,F)\int_{\Sigma} P^{\widehat{A}}(\mathrm{d}\lambda) .
\end{equation}
Equation (\ref{newoperator}) defines a linear, bounded, positive operator which depends not only on $F$ but also on $P$. It is then natural to assume that, for every pure state $P$ and generalized observable $A_0 \in {\mathscr O}_{0}$, a mapping $p_{P,A_{0}}^{d}:\Xi_{A_{0}}\longrightarrow[0,1]$ exists such that
\begin{equation} \label{POV_lebesgue}
T_{P,A_{0}}(\Sigma) =\int_{\Sigma}{p}_{P,A_{0}}^{d}(\lambda) P^{\widehat{A}}(\mathrm{d}\lambda) \quad (a_0 \notin \Sigma) .
\end{equation}
Hence,
\begin{equation}\label{boggi}
p^{t}(P,F)=Tr[\rho_{P}\int_{\Sigma}{p}_{P,A_{0}}^{d}(\lambda) P^{\widehat{A}}(\mathrm{d}\lambda)]
\end{equation}
and
\begin{equation}
p^{d}(P,F)=\frac{Tr[\rho_{P}\int_{\Sigma}{p}_{P,A_{0}}^{d}(\lambda) P^{\widehat{A}}(\mathrm{d}\lambda)]}{Tr[\rho_{P}\int_{\Sigma} P^{\widehat{A}}(\mathrm{d}\lambda)]} .
\end{equation}
Therefore, as far as $p^{t}(S,F)$ is concerned, the pure state $P$ can still be represented by $\rho_{P}$. The representation of the physical property $F$ varies instead with $P$, so that $F$ is represented by the family $\{ T_{P,A_{0}}(\Sigma) \}_{P \in {\mathscr P}}$.

Let us consider now a physical property $F=(A_0, \Sigma)\in {\mathscr F}_{0}\setminus{\mathscr F}$ (hence $a_0\in \Sigma$). By using Eqs. (\ref{f_bar_c}), (\ref{POV_lebesgue}) and (\ref{boggi}) we obtain
\begin{equation} \label{prob_X0_W}
p^{t}(P,F)=1-Tr[\rho_{P}T_{P,A_{0}}(\Xi_{A_0}\setminus\Sigma)]=Tr[\rho_{P}T_{P,A_{0}}(\Sigma)]
\end{equation}
with
\begin{equation} \label{POV_lebesgue_0}
T_{P,A_{0}}(\Sigma) =I-\int_{\Xi_{A_{0}}\setminus\Sigma}{p}_{P,A_{0}}^{d}(\lambda) P^{\widehat{A}}(\mathrm{d}\lambda) \quad (a_0 \in \Sigma) .
\end{equation}
where $I$ is the identity operator on $\mathscr H$.

Putting together Eqs. (\ref{prob_X_W}) and (\ref{POV_lebesgue_0}) we obtain that the mapping $\widehat{p^{t}}$ defined by Eq. (\ref{pt}) is a probability measure on $\Xi_{A_0}$, as required in Sect. \ref{esrsub1}.

For every pure state $P$ represented by $\rho_P$ we can thus introduce a (commutative) positive operator valued (POV) measure   
\begin{equation} \label{math_rep_gen_obs}
T_{P,A_{0}}: \Sigma \in \mathbb{B}(\Re) \longmapsto T_{P,A_{0}}(\Sigma) \in {\mathscr B}({\mathscr H}) ,
\end{equation}
where ${\mathscr B}({\mathscr H})$ is the set of all bounded operators on $\mathscr H$, defined by Eqs. (\ref{POV_lebesgue}) and (\ref{POV_lebesgue_0}). Moreover, for every Borel set $\Sigma\subset \Xi_{A_0}$, the family
\begin{equation} \label{family_gen_bs}
{\mathcal T}_{A_{0}}=\left \{ T_{P,A_{0}}\right \}_{P \in \mathscr P}
\end{equation}
allows one to calculate, via Eqs. (\ref{prob_X_W}) or (\ref{prob_X0_W}), the overall probability that the outcome of a measurement of the generalized observable $A_0$ on an individual object $\alpha$ in the state $P$ belongs to $\Sigma$. Hence we can assume that $A_0$ is represented by the family ${\mathcal T}_{A_0}$ as far as the overall probability is concerned.

\vspace{1mm}
\noindent
Putting together the results in (i) and (ii) we conclude that, whenever only pure states are considered, the overall and the conditional on detection probabilities can be calculated by using the representation of pure states supplied by QM. The mathematical representation of a physical property $F=(A_0,\Sigma)\in \mathscr F$ is instead provided by the pair $(P^{\widehat{A}}(\Sigma), \{ T_{P}^{\widehat{A}}(\Sigma)\}_{P \in \mathscr P})$. The first element of the pair coincides with the representation of $F$ supplied by QM and must be used to calculate $p(P,F)$. The second element of the pair is specific of the ESR model and must be used to calculate $p^{t}(P,F)$. Analogously, the representation of a generalized observable $A_0 \in {\mathcal O}_{0}$ is provided by the pair
$(\widehat{A},{\mathcal T}_{A_0})$. The first element of the pair coincides with the representation supplied by QM of the observable $A \in \mathscr O$ from which $A_0$ is obtained. The second element of the pair is specific of the ESR model.

One can now stem from the mathematical representations reported above to discuss how overall and conditional on detection probability can be calculated in the case of mixtures. For the sake of brevity we do not discuss the details of this treatment here, and only report the results that have been obtained by two of us \cite{gs12}.

Let us begin with a preliminary remark. We have mentioned in Sect. \ref{sub1} the distinction between proper and improper mixtures. This distinction is often ignored by physicists because the two kinds of mixtures are represented by the same mathematical entities (density operators) in QM. But several scholars have pointed out that proper and improper mixtures can be empirically distinguished \cite{tb05}, which implies that some physical information is lost in the mathematical representation. This is the deep reason of the problems that arise in QM when one tries to provide a physical interpretation of the coefficients that occur in the decompositions of mixtures in terms of pure states. These problems are avoided in the ESR model, which takes into account the differences in the empirical interpretations (or \emph{operational definitions}) of the two kinds of mixtures, supplying different mathematical representations of them.

Firstly, let us consider a proper mixture $M \in \mathscr M$ of the pure states $P_1, P_2, \ldots$, with probabilities $p_1, p_2, \ldots$, respectively. Then, $M$ is represented in the ESR model by a family of pairs  $\{ ( \rho_{M}(F), p^{d}(M,F) ) \}_{F \in {\mathcal F}}$. For every $F=(A_0,\Sigma) \in \mathscr F$, $\rho_{M}(F)$ is a density operator given by
\begin{equation} 
\rho_{M}(F)=\frac{\sum_{j}p_{j}\frac{Tr[\rho_{P_{j}}T_{P_{j},A_{0}}(\Sigma)]}{Tr[\rho_{P_{j}}P^{\widehat{A}}(\Sigma)]}\rho_{P_{j}}}{\sum_{j}p_{j}\frac{Tr[\rho_{P_{j}}T_{P_{j},A_{0}}(\Sigma)]}{Tr[\rho_{P_{j}}P^{\widehat{A}}(\Sigma)]}}
\end{equation}
and $p^{d}(M,F)$ is a detection probability given by
\begin{equation} 
p^{d}(M,F)=\sum_{j}p_{j}p^{d}(P_{j},F).
\end{equation}
The conditional on detection and the overall probability are given by
\begin{equation} \label{uk}
p(M,F)=Tr [ \rho_{M}(F) P^{\widehat{A}}(\Sigma)]
\end{equation}
and 
\begin{equation} \label{prob_M}
p^{t}(M,F)=Tr [\rho_{M}(F) T_{M,A_{0}}(\Sigma)],
\end{equation}
respectively, with
\begin{equation} \label{POV_lebesgue_M}
T_{M,A_{0}}(\Sigma)=p^{d}(M,F) P^{\widehat{A}}(\Sigma).
 \end{equation}
 
Secondly, let us consider an improper mixture $N \in \mathscr N$. Then, $N$ can be represented by the same density operator $\rho_{N}$ that represents it in QM, and the conditional on detection probability is given by
\begin{equation} \label{prob_X_N}
p(N,F)=Tr [\rho_{N} P^{\widehat{A}}(\Sigma)].
\end{equation}
Because of Eq. (\ref{prob_X_N}) assumption AX 3 can be extended to improper mixtures. Moreover, a linear, bounded, positive operator $T_{N,A_{0}}(\Sigma)$ can be introduced as in the case of pure states, whose expression is given by Eqs. (\ref{POV_lebesgue}) and (\ref{POV_lebesgue_0}), with $N$ in place of $P$. The overall probability is then given by
\begin{equation} \label{prob_X_NOverall}
p^{t}(N,F)=Tr [\rho_{N} T_{N,A_{0}}(\Sigma)].
\end{equation}
Hence the set of improper mixtures can be considered as an extension of the set of pure states, and improper mixtures as \emph{generalized pure states} \cite{gs12,a99}.

Coming to physical properties, Eqs. (\ref{uk})--(\ref{prob_X_NOverall}) show that the mathematical representation of a physical property $F=(A_0,\Sigma)\in \mathscr F$ which holds in the case of pure states can be extended to mixtures. To be precise, the property $F$ is represented by the pair $(P^{\widehat{A}}(\Sigma), \{ T_{S,A_{0}}(\Sigma)\}_{S \in \mathscr S})$.

As we have seen above, the difference between the mathematical representations of proper and improper mixtures corresponds to the empirical difference between the two kinds of mixtures, which is epistemologically satisfactory and avoids the interpretative problems that arise in QM. Moreover the difference between the quantum description and the ESR model description of proper mixtures implies that possible experiments aiming to check which of the two theories provides correct predictions can be contrived \cite{gs12}. 

\subsection{Idealized measurements\label{esrsub5}}
The representations worked out in Sect. \ref{esrsub4} suggest how to modify L\"{u}ders' postulate of QM (Sect. \ref{sub3}) to select a class of measurements analogous to the first kind, ideal measurements of QM. To be precise, let $\alpha$ be an individual object in a state $S \in {\mathscr P}\cup{\mathscr N}$ (that is, $S$ is either a pure state or an improper mixture), represented by the density operator $\rho_S$. Then, we assume that, for every physical property $F=(A_0,\Sigma)\in {\mathscr F}_{0}$, a (nondestructive, exact) \emph{idealized} measurement exists that transforms $S$ into the final state $S_F$ represented by the density operator 

\begin{equation} \label{genpost_dis_W}
\rho_{S_{F}}=\frac{T_{S,A_{0}}(\Sigma) \rho_{S} T_{S,A_{0}}(\Sigma)}{Tr[T_{S,A_{0}}(\Sigma) \rho_{S} T_{S,A_{0}}(\Sigma)]}
\end{equation}
if the \emph{yes} result is obtained. In analogy with QM we call the rule expressed by Eq. (\ref{genpost_dis_W}) \emph{generalized L\"{u}ders' postulate} (GLP) in the following.

We stress that Eq. (\ref{genpost_dis_W}) does not apply if an idealized measurement is performed on an individual object in a state $M \in \mathscr M$ (proper mixture). However, the state transformation induced in this case can be deduced from Eq. (\ref{genpost_dis_W}). Its expression is rather complicate \cite{gs12,gs10} and we do not report it here for the sake of brevity. 

Finally, let us note that we will often refer  to the special case of a pure state $P$ and a discrete generalized observable $A_0$ in the following. Therefore let us discuss how our general formulas particularize in this specific case. If $A_0$ is obtained from a discrete observable $A$ of QM whose set of possible outcomes is $\Xi= \{ a_1, a_2, \ldots, a_W \}$, with $W$ finite or infinite, the set of possible outcomes of $A_0$ is $\Xi_{A_0}= \{ a_0, a_1, a_2, \ldots, a_W\}$. Let us denote by $P_{1}^{\widehat{A}}, P_{2}^{\widehat{A}}, \ldots,  P_{W}^{\widehat{A}}$ the (orthogonal) projection operators associated with $a_1, a_2, \ldots, a_W$, respectively, by the spectral decomposition of $\widehat{A}$. Then we get from Eqs. (\ref{POV_lebesgue}) and (\ref{POV_lebesgue_0})
\begin{equation} \label{caso_particolare_X}
T_{P,A_{0}}(\Sigma) = \left \{
\begin{array}{cll} 
\sum_{n | a_n \in \Sigma }p_{P, A_{0}}^{d}(a_n) P_{n}^{\widehat{A}} & & \textrm{if} \ a_0 \notin \Sigma  \\
I -\sum_{n | a_n \in \Xi_{A_0} \setminus \Sigma} p_{P,A_{0}}^{d}(a_n) P_{n}^{\widehat{A}} & & \textrm{if} \ a_0 \in \Sigma
\end{array}
\right. .
\end{equation}
Let $\Sigma=\{ a_k \}$, with $k \in \{ 1,2,\ldots,W\}$. Then Eq. (\ref{caso_particolare_X}) yields
\begin{equation} \label{caso_particolare_uno}
T_{P,A_{0}}(\{ a_k \}) = \left \{
\begin{array}{cll} 
p_{P,A_{0}}^{d}(a_k)P_{k}^{\widehat{A}} & & \textrm{if} \ k \ne 0  \\
\sum_{n=1}^{W} (1-p_{P,A_{0}}^{d}(a_n))P_{n}^{\widehat{A}} & & \textrm{if} \ k=0
\end{array}
\right. .
\end{equation}
If we put $F_k=(A_0, \{ a_k \})$, Eqs. (\ref{prob_X_W}) and (\ref{prob_X0_W}) yield
\begin{equation} \label{caso_particolare_n}
p^{t}(P,F_k)=\left \{
\begin{array}{cll}
Tr[ \rho_{P} p_{P, A_{0}}^{d}(a_k)  P_{k}^{\widehat{A}}] & &  \textrm{if} \ k \ne 0 \\
Tr[ \rho_{P} \sum_{n=1}^{W} (1-p_{P, A_{0}}^{d}(a_n)) P_{n}^{\widehat{A}}] & & \textrm{if} \ k=0
\end{array}
\right. .
\end{equation}
Whenever the property $F_k$ is measured and the yes outcome is obtained, Eq. (\ref{genpost_dis_W}) yields
\begin{equation} \label{gpp_improper}
\rho_{P_{F_k}} = \left \{
\begin{array}{cll} 
\frac{P_{k}^{\widehat{A}}\rho_{P} P_{k}^{\widehat{A}}}{Tr [P_{k}^{\widehat{A}}\rho_{P} P_{k}^{\widehat{A}}]} & & \textrm{if} \ k \ne 0  \\
\frac{\sum_{m,n=1}^{W} (1-p_{P, A_{0}}^{d}(a_m))(1-p_{P,A_{0}}^{d}(a_n))P_{m}^{\widehat{A}}\rho_{P} P_{n}^{\widehat{A}}}{Tr \Big [ \sum_{m,n=1}^{W} (1-p_{P,A_{0}}^{d}(a_m))(1-p_{P,A_{0}}^{d}(a_n))P_{m}^{\widehat{A}}\rho_{P} P_{n}^{\widehat{A}} \Big ]} & & \textrm{if} \ k=0
\end{array}
\right. .
\end{equation} 
For the sake of simplicity and intuitivity we will use sometimes the representation of pure states by means of unit vectors of $\mathscr H$ in the following. We therefore observe that, if $P \in {\mathscr P}$ is represented by the unit vector $|\psi\rangle \in {\mathscr V}$, the state $P_{F_{k}}$ after a measurement of $F_k$ which yields result \emph{yes} is represented by the unit vector
\begin{equation} \label{caso_particolare_psi_uno}
|\psi_{F_k}\rangle = \left \{
\begin{array}{cll} 
\frac{P_{k}^{\widehat{A}}|\psi\rangle}{\sqrt{\langle\psi|P_{k}^{\widehat{A}}|\psi\rangle}} & & \textrm{if} \ k \ne 0  \\
\frac{\sum_{n=1}^{W} (1-p_{P,A_{0}}^{d}(a_n))P_{n}^{\widehat{A}}|\psi\rangle}{\sqrt{\sum_{n=1}^{W} (1-p_{P,A_{0}}^{d}(a_n))^{2} \Vert P_{n}^{\widehat{A}}|\psi\rangle \Vert^{2}}} & & \textrm{if} \ k=0
\end{array}
\right. 
\end{equation} 
If $k \ne 0$, Eq. (\ref{caso_particolare_psi_uno}) reproduces the standard form of the \emph{projection postulate} that can be found in the manuals of QM. If $k=0$, it shows that the initial state can be modified by the measurement even if the individual object is not detected, though this does not occur for special classes of generalized observables \cite{sg08}.

\section{A hidden variables theory of the measurement process in the ESR model\label{hv}}
We intend to supply in this section a hidden variables theory of the measurement process in the ESR model  which shows that the ESR model can be considered as an \emph{objective} theory, at variance with QM, if suitable conditions on the detection probability are satisfied. To this end, we add a set of theoretical \emph{microscopic} entities to the theoretical entities (that we call \emph{macroscopic} in the following because of the empirical interpretation in Sect. \ref{esrsub2}) introduced in Sect. \ref{esrsub1} to describe the physical system $\Omega$. Intuitively, the link between the macroscopic and the microscopic entities is established by the set $\mathscr U$ of all individual objects introduced in Sect. \ref{esrsub2}. Individual objects are supposed indeed to have microscopic properties which determine the outcomes of the measurements of (macroscopic) physical properties and the probabilities introduced in Sect. \ref{esrsub1} (we stress, however, that no model of individual objects as classical or semi-classical particles is presupposed in the ESR model). This intuitive idea can be implemented as follows.

\subsection{Microscopic properties and states\label{submicro1}}
We assume that a physical system $\Omega$ is characterized by a set ${\mathscr F}_{\mu}$ of \emph{microscopic properties} at a microscopic level. The elements of ${\mathscr F}_{\mu}$ are the \emph{hidden variables} of the models. Each microscopic property $f \in {\mathscr F}_{\mu}$ is a mapping $f: \alpha \in {\mathscr U} \longrightarrow f(\alpha) \in \{0, 1\}$. Hence, for every individual object $\alpha \in {\mathscr U}$, the set ${\mathscr F}_{\mu}$ is partitioned in two subsets, the subset $S_{\mu}= \{ f\in {\mathscr F}_{\mu} | \ f(\alpha)=1 \}$ of microscopic properties that are \emph{possessed} by $\alpha$, and the subset ${\mathscr F}_{\mu}\setminus {S}_{\mu}=\{ f\in {\mathscr F}_{\mu} | \ f(\alpha)=0 \}$ of microscopic properties that are \emph{not possessed} by $\alpha$ (note that the terms ``possessed'' and ``not possessed'' have no empirical interpretation at this stage: hence they do not refer to any measurement procedure). The
set $S_\mu$ is then called the \emph{microscopic state} of $\alpha$, and one briefly says that \emph{$\alpha$ is in the microscopic state $S_\mu$}. Furthermore, the set of all individual objects in
the microscopic state $S_\mu$ (that is, the set of all individual objects which possess
all the microscopic properties that belong to $S_\mu$, and only those) is called the \emph{extension} of $S_{\mu}$ and is denoted by $\textrm{ext} S_\mu$, while the set of all possible microscopic states of $\Omega$ is denoted by ${\mathscr S}_{\mu}$. It is then apparent that the family $\{ \textrm{ext} S_{\mu} \}_{S_{\mu} \in {\mathscr S}_{\mu}}$ is a partition of $\mathscr U$.\footnote{Note that the family $\{ \textrm{ext} S_{\mu} \cap \textrm{ext} S \}_{S_{\mu} \in {\mathscr S}_{\mu}, S \in {\mathscr S}}$ is a further partition of $\mathscr U$, some elements of which may be void.} The basic link between microscopic and macroscopic entities is now established by assuming that a bijective mapping $\varphi: f\in {\mathscr F}_{\mu} \longrightarrow F \in  {\mathscr F} \subset{\mathscr F}_{0}$ exists which makes every microscopic property correspond to a physical property of the subset $\mathscr F$ introduced in Sect. \ref{esrsub1}. Because of this assumption one can associate each microscopic state $S_{\mu}$ with a set $ \{F \in {\mathscr F} \ | \ \varphi^{-1}(F) \in S_{\mu} \}$ of physical properties or, equivalently, with a set $\{E \in {\mathscr E} \ | \ \varphi^{-1}(g^{-1}(E)) \in S_{\mu} \}$ of quantum properties.

The result of an (exact) measurement of a (macroscopic) physical property $F \in \mathscr F$ on an individual object $\alpha$ in the state $S \in \mathscr S$ is explained at a microscopic level as follows. The set of all microscopic properties possessed by $\alpha$, that is, the microscopic state $S_{\mu}$ of $\alpha$, induces a probability that the registering device react or, equivalently, that $\alpha$ be detected. Let $f=\varphi^{-1}(F)$. Then the measurement of $F$ yields the outcome \emph{yes} if $\alpha$ is detected and possesses $f$ (we say that $\alpha$ \emph{displays} $F$ in this case, see Sect. \ref{sub2}), while it yields the outcome \emph{no} if $\alpha$ is not detected or does not possess $f$ (we say that $\alpha$ displays the complementary property $F^{c}$ of $F$ in this case). The result of the measurement of a physical property $F \in {\mathscr F}_{0}\setminus{\mathscr F}$ is then explained by considering $F^{c}$ in place of $F$ and exchanging \emph{yes} and {\it no}.

The explanation above implies that, whenever $\alpha$ is detected, it displays the physical property $F$ iff $f \in S_{\mu}$. We are thus led to introduce the following probabilities.

$p^{d}(S_{\mu},F)$: the \emph{microscopic detection probability}, that is, the probability that an individual object $\alpha$ in the microscopic state $S_\mu$ be detected when $F$ is measured on it.

$p(S_{\mu},F)$: the \emph{microscopic conditional on detection probability}, that is, the probability that an individual object $\alpha$ in the microscopic state $S_\mu$ display $F$ when $F$ is measured on it and $\alpha$ is
detected (which is either 0 or 1 since $\alpha$ either possesses $f = \varphi^{-1}(F)$ or not, because either $f = \varphi^{-1}(F)\in S_{\mu}$ or $f = \varphi^{-1}(F)\notin S_{\mu}$).

$p^{t}(S_{\mu},F)$: the \emph{microscopic overall probability}, that is, the probability that an individual object $\alpha$ in the microscopic state $S_\mu$ display $F$ when $F$ is measured on it.

Hence, we get
\begin{equation}\label{microprob}
p^{t}(S_{\mu},F)=p^{d}(S_{\mu},F)p(S_{\mu},F)
\end{equation}
Equation (\ref{microprob}) is purely theoretical, because one can never directly prepare an individual object in the microscopic state $S_{\mu}$. Indeed one can only choose a device $\pi \in \Pi$ and then prepare $\alpha$ by means of $\pi$, so that $\alpha$ is in the (macroscopic) state $S \in \mathscr S$ empirically interpreted on $[\pi]_{\equiv}$. For every $\alpha$ in the state $S$ let
us therefore introduce a further conditional probability, as follows.

$p(S_{\mu}|S)$: the conditional probability that an individual object $\alpha$ be in the microscopic state $S_\mu$ whenever it is in the state $S$.

We can thus associate a subset ${\mathscr S}_{\mu|S}$ of microscopic states with every macroscopic state $S \in \mathscr S$
\begin{equation}
{\mathscr S}_{\mu|S}=\{ S_{\mu} \in {\mathscr S}_{\mu} \ | \ p(S_{\mu}|S) \ne 0 \}.\footnote{It is then easy to show that ${\mathscr S}_{\mu|S}=\{ S_{\mu} \in {\mathscr S}_{\mu} \ | \ \textrm{ext}S_{\mu} \cap \textrm{ext}S \ne 0 \}$.}
\end{equation}

The joint probability that an individual object $\alpha$ in the state $S$ be in the microscopic state $S_{\mu} \in{\mathscr S}_{\mu|S}$  and display $F$ when $F$ is measured on it is then given by $p(S_{\mu}|S) p^{t}(S_{\mu},F)$. Hence the overall probability $p^{t}(S,F)$ that an individual object $\alpha$ in the state $S$  display $F$ when $F$ is measured on it is\footnote{For the sake of simplicity, we consider here only the discrete case. Note that the sum can be extended to all microscopic states in ${\mathscr S}_{\mu}$, because $p(S_{\mu}|S)=0$ if $S_{\mu} \notin {\mathscr S}_{\mu|S}$.}
\begin{equation}\label{quellat}
p^{t}(S,F)=\sum_{S_{\mu} \in {\mathscr S}_{\mu|S}} p(S_{\mu}|S) p^{t}(S_{\mu},F).
\end{equation}
Moreover, the detection probability $p^{d}(S,F)$ that an individual object $\alpha$ in the state $S$ be detected when $F$ is measured on it is given by
\begin{equation}\label{quella}
p^{d}(S,F)=\sum_{S_{\mu} \in {\mathscr S}_{\mu|S}} p(S_{\mu}|S) p^{d}(S_{\mu},F).
\end{equation}
Let us define now
\begin{equation}
p(S,F)=\frac{\sum_{S_{\mu} \in {\mathscr S}_{\mu|S}} p(S_{\mu}|S) p^{t}(S_{\mu},F)}{\sum_{S_{\mu} \in {\mathscr S}_{\mu|S}} p(S_{\mu}|S) p^{d}(S_{\mu},F)}.
\end{equation}
Then, for every $S \in {\mathscr S}$ and $F \in {\mathscr F}$, we obtain
\begin{equation}\label{formuladipartenzaprop}
p^{t}(S,F)=p^{d}(S,F)p(S,F).
\end{equation}
Equation (\ref{formuladipartenzaprop}) coincides with Eq. (\ref{ax1}). Hence, it justifies it in terms of the hidden variables (microscopic properties) that have been introduced. The crucial feature of this derivation is that no--detection is caused only by the microscopic properties possessed by $\alpha$. Indeed, these properties determine the probability $p^{d}(S_{\mu},F)$, while the conditional probability $p(S_{\mu}|S)$ depends only on $S_{\mu}$ and $S$. Hence, Eq. (\ref{quella}) implies that $p^{d}(S,F)$ is noncontextual, in the sense that it is determined only by the microscopic properties of the individual objects in $\textrm{ext}S$, as stated, and neither occurs because of flaws or lack of efficiency of the apparatus measuring $F$ nor it depends on the physical context in which $F$ is measured (we stress however that $p^{d}(S,F)$ depends on $F$: if $F=(A_0,\Sigma)$, $p^{d}(S,F)$, generally, is not fixed for a given generalized observable $A_0$ and depends on $\Sigma$).

To complete our discussion it remains to consider the measurement of a property $F\in {\mathscr F}_{0}\setminus{\mathscr F}$. To this end let us still denote by $p^{t}(S_\mu,F)$ the overall probability that an individual object $\alpha$ in the microscopic state $S_{\mu} \in {\mathscr S}_{\mu|S}$ display $F$ when $F$ is measured on it and recall that $F^{c}=(A_0, \Xi_{A_0} \setminus \Sigma)$. Then, we introduce the physically reasonable assumption that, for every $S_{\mu} \in {\mathscr S}_{\mu|S}$,
\begin{equation}\label{reasonassumpt}
p^{t}(S_\mu,F)=1-p^{t}(S_\mu,F^{c}),
\end{equation}
Equation (\ref{reasonassumpt}) yields $p^{t}(S_\mu,F)$ in terms of the overall probability that $\alpha$ display $F^c$ when $F^c$ is measured in place of $F$, which is given by Eq. (\ref{microprob}), with $F^c$ in place of $F$. Then, reasoning as in the case of Eq. (\ref{quellat}), we get
\begin{eqnarray}
p^{t}(S,F)&=&\sum_{S_{\mu} \in {\mathscr S}_{\mu|S}} p(S_{\mu}|S) p^{t}(S_{\mu},F)=\sum_{S_{\mu} \in {\mathscr S}_{\mu|S}} p(S_{\mu}|S)(1-p^{t}(S_\mu,F^{c})) \nonumber \\
&=&\sum_{S_{\mu} \in {\mathscr S}_{\mu|S}} p(S_{\mu}|S)-\sum_{S_{\mu} \in {\mathscr S}_{\mu|S}} p(S_{\mu}|S) p^{t}(S_\mu,F^{c}).
\end{eqnarray}
Bearing in mind that $p^{t}(S,F^{c})=\sum_{S_{\mu} \in {\mathscr S}_{\mu|S}} p(S_{\mu}|S) p^{t}(S_{\mu},F^{c})$, because of Eq. (\ref{quellat}), and that $\sum_{S_{\mu} \in {\mathscr S}_{\mu|S}} p(S_{\mu}|S)=1$, we obtain the following equation, which holds for every $S \in \mathscr S$ and $F \in {\mathscr F}_{0}\setminus {\mathscr F}$,
\begin{equation}\label{f_bar_cprop}
p^{t}(S,F)=1-p^{t}(S,F^{c})
\end{equation}
or, equivalently
\begin{equation}
p^{t}(S,F)=1-p^{d}(S,F^{c})p(S,F^{c}).
\end{equation}
Equation (\ref{f_bar_cprop}) coincides with Eq. (\ref{f_bar_c}). Hence, also this equation is justified in terms of the hidden variables that have been introduced. Moreover, also in this case $p^{d}(S,F^{c})$ is noncontextual in the sense specified above.

\subsection{The objectivity issue in the ESR model\label{submicro2}}
The hidden variables theory for the measurement process that has been constructed in Sect. \ref{submicro1} can be specialized in various ways by introducing different assumptions on $p^{d}(S_{\mu},F)$ and $p(S_{\mu},F)$. The simplest possible assumption is that $p^{d}(S_{\mu},F)\in \{0,1\}$ (which does not imply $p^{d}(S,F)\in \{0,1\}$), that is, intuitively, that the physical properties possessed by an individual object $\alpha \in {\rm ext}S$ determine whether $\alpha$ is detected or not when $F$ is measured. We call the hidden variables theory {\it deterministic} in this case. More generally, we can assume that $p^{d}(S_{\mu},F)$ admits an epistemic interpretation in terms of further unknown features of the individual objects in the state $S_\mu$, which can be formalized by introducing an additional hidden variable besides microscopic properties  \cite{gs08}. The hidden variables theory thus provides a description of the measurement process in the ESR model according to which the outcome of a measurement of a physical property $F$ is prefixed for every measurement context (condition VD in Sect. \ref{objectivity}) and independent of the measurement context (condition NC in Sect. \ref{objectivity}). Based on this description, AX 1 and AX 2 in Sect. \ref{esrsub3} are recovered. Nevertheless, we cannot incorporate such description within the ESR model and conclude that this model is an objective theory without due care. Indeed, it can be incompatible with AX 3 if the detection probabiltiy is not suitably chosen. Suppose, for example, that $p^{d}(S,F)$ is identically 1. In this case the ESR model coincides with QM, hence it is nonobjective.

In our former presentations of the ESR model the hidden variables theory and the macroscopic part of the ESR model were intertwined, and the whole model was assumed to be objective. In this perspective AX 3 introduced {\it consistency conditions} on the detection probability. But we could not state a general form for such conditions, because we have as yet no theory which allows us to predict the values of the detection probability. We therefore studied several special cases and singled out the conditions to be fulfilled to make the objective description provided by the ESR model consistent. In particular, we considered the original Bell inequality and the Clauser--Horne--Shimony--Holt (CHSH) inequality. Both are obtained by assuming ``local realism'' ({\it i.e.}, VD and NC at a distance, or {\it locality}, in our present terms, see Sects. \ref{intro} and \ref{objectivity}) and then dealing with the Bohm variant of the Einstein--Podolsky--Rosen (EPR) thought experiment. It is well known that the foregoing inequalities conflict with the quantum description of the experiment, which is usually maintained to be a proof of the unavoidable nonlocality (hence, contextuality) of QM. We proved that the Bell and CHSH inequalities must be modified according to the ESR model, inserting in them several values of the detection probability \cite{gp13,s07,gs08,gs10,gs10b}. The modified inequalities hold together with the quantum inequalities in the ESR model. This can be explained by observing that the former refer to the set of all individual objects that are produced, while the latter refer to the set of all individual objects that are detected when exact measurements are performed. Elementary estimations under very restrictive conditions then show that no contradiction occurs if the efficiency of the detectors is lower than 0.8165 in the case of the Bell inequality, and 0.841 i the case of the CHSH inequality \cite{s07,gs08,gs10b}. If the values of the detection probabilities are considered as unknown parameters to be determined empirically, the results of suitable experiments can confirm or falsify the ESR model in such specific cases.

We have also discussed the Greenberger--Horne--Zeilinger (GHZ) experiment \cite{ghsz82} that is usually maintained to prove the nonlocality of QM without resorting to inequalities. We have proven that the ``toy'' hidden variables models proposed by  Szab\'o and Fine \cite{sf02} to supply  a local (and noncontextual) explanation of the results predicted by QM can be obtained as special cases of the hidden variables theory incorporated in the ESR model, with suitable choices of the values of the detection probability \cite{gps13}.

We have implicitly propounded, however, a different perspective on the previous sections of the present paper. We have avoided intertwining the macroscopic and the microscopic part of the ESR model, considering the latter part as a hidden variable theory of the measurement process introduced in the former. In this view the ESR model may be objective or not, depending on the values of the detection probability. The consistency conditions mentioned above can then be reinterpreted as {\it demarcation conditions}, which establish the border that must not be trespassed if the description provided by the ESR model has to be objective.

All the foregoing results can be restated in our new perspective. In the case of the Bell and CHSH inequalities, the violation of the limits reported above would not imply that the ESR model is falsified, but, rather, that it does not provide an objective description of the physical system. In the case of the GHZ experiment the recovery of Szab\'o and Fine local models shows, by means of examples, that the ESR model can actually supply objective descriptions of composite physical systems if the values of the detection probability are suitably chosen.

\section{Time evolution in the ESR model\label{evolution_meas}}
Our presentation of the ESR model in Sect. \ref{modello} did not explicitly discuss time evolution, but it implicitly introduced changes of states with time when considering idealized measurements. If we assume that our generalization and reinterpretation of QM can be applied to composite systems made up of a physical system and by a macroscopic registering device (see footnote \ref{notemeasurement}), we can consider this specific case as a guide for contriving a general description of time evolution in the ESR model. We therefore provide a simple measurement scheme in the next section, describing an idealized measurement as an interaction between physical systems. The obtained results constitute a basis for discussing whether linear unitary evolution of the composite system may occur, stating some general assumptions on time evolution and partially justifying the hypotheses on idealized measurements introduced in Sect. \ref{esrsub5}. For the sake of intuitivity, we will firstly consider pure states and discrete generalized observables only.

\subsection{Time evolution induced by measurements\label{subevo1}}
Let $\Omega$ be a physical system associated with the Hilbert space $\mathscr H$, and let  $\Omega^{M}$ be an apparatus (hence a macroscopic physical system) which performs an idealized measurement of a discrete generalized observable $A_0$ of $\Omega$ obtained from a discrete quantum observable $A$ of $\Omega$. By using the symbols introduced in Sect. \ref{esrsub5}, the possible values $a_0, a_1, a_2, \ldots, a_W$ of $A_0$ (with $W$ finite or infinite) then bijectively correspond to the positions $v(a_0)$, $v(a_1)$, $v(a_2)$, \ldots, $v(a_W)$, respectively, of a pointer of $\Omega^{M}$. Let us maintain that the ESR model applies also to macroscopic physical systems and to any composite system (thus implicitly claiming the universality of the ESR model, see footnote \ref{notemeasurement}). Hence $\Omega^{M}$ is associated with the Hilbert space ${\mathscr H}^{M}$, and the positions ({\it outcomes}) $v(a_0)$, $v(a_1)$, $v(a_2)$, \ldots, $v(a_W)$ correspond to states $S_0^{M}, S_1^{M}, S_2^{M}, \ldots, S_{W}^{M}$ of $\Omega_M$, respectively. In our simplified scheme these states are assumed to be pure: hence, they are represented by unit vectors $|a_{0}^{M}\rangle, |a_{1}^{M}\rangle, |a_{2}^{M}\rangle$, \ldots, $|a_{W}^{M}\rangle$ of ${\mathscr H}^{M}$, respectively. We denote by ${\mathscr G}^{M}$ the subspace $\langle \{  |a_{0}^{M}\rangle, |a_{1}^{M}\rangle, |a_{2}^{M}\rangle$, \ldots, $|a_{W}^{M}\rangle \} \rangle$ of ${\mathscr H}^{M}$ generated by these vectors in the following.

Coming to $\Omega$, we recall from Sect. \ref{esrsub4} that the mathematical representation of the generalized observable $A_0$ is provided by the pair $(\widehat{A},{\mathcal T}_{A_0})$, with $\widehat{A}$ the self-adjoint operator representing the observable $A$ in QM. The values $a_1$, $a_2$, \ldots, $a_W$ then are eigenvalues of $\widehat{A}$. We denote by ${\mathscr S}_{1}^{\widehat{A}}$, ${\mathscr S}_{2}^{\widehat{A}}$, \ldots, ${\mathscr S}_{W}^{\widehat{A}}$ the subspaces associated with
$a_1$, $a_2$, \ldots, $a_W$, respectively, and put $g_n={\rm dim} {\mathscr S}_{n}^{\widehat{A}}$ ($n=1,2,\ldots,W$). For every 
${\mathscr S}_{n}^{\widehat{A}}$, we introduce an orthonormal basis $\{|a_{n}^{\mu}\rangle \}_{\mu=1,2,\ldots,g_n}$ of vectors of ${\mathscr S}_{n}^{\widehat{A}}$, so that $\{|a_{n}^{\mu}\rangle \}_{n=1,2,\ldots,W;\mu=1,2,\ldots,g_n}$ is an orthonormal basis on ${\mathscr H}$.\footnote{We observe that the mapping $\tau:{\mathscr S}_{n}^{\widehat{A}} \in \{{\mathscr S}_{1}^{\widehat{A}}$, ${\mathscr S}_{2}^{\widehat{A}}$, \ldots, ${\mathscr S}_{W}^{\widehat{A}}\}\longrightarrow |a_{n}^{\mu}\rangle \in \{|a_{1}^{\mu}\rangle, |a_{2}^{\mu}\rangle,\ldots,|a_{W}^{\mu}\rangle \}$ canonically induces a homomorphism of  ${\mathscr H}$ onto the proper subspace of  ${\mathscr G}^{M}$ generated by the set $\{ |a_{1}^{M}\rangle, |a_{2}^{M}\rangle$, \ldots, $|a_{W}^{M}\rangle\}$ of unit vectors of ${\mathscr G}^{M}$.} The rules for calculating overall probabilities and state transformations in the case of idealized measurements are then given by Eqs. (\ref{caso_particolare_n}) and (\ref{caso_particolare_psi_uno}), respectively (with $P_{k}^{\widehat{A}}$ and $P_{n}^{\widehat{A}}$ the orthogonal projection operators whose ranges are ${\mathscr S}_{k}^{\widehat{A}}$ and ${\mathscr S}_{n}^{\widehat{A}}$, respectively).

By using the symbols introduced above and in Sect. \ref{esrsub5}, we can  describe an idealized measurement by considering the composite physical system $(\Omega,\Omega^{M})$.

According to the ESR model, pure states and improper mixtures of $(\Omega,\Omega^{M})$  can be represented as in QM. Hence, $(\Omega,\Omega^{M})$ can be associated with the Hilbert space $\mathscr H\otimes{\mathscr H}^{M}$, and Eq. (\ref{caso_particolare_psi_uno}) suggests characterizing idealized measurements by means of the following axiom.

\vspace{1mm}
\noindent
\textbf{AXM.} Let an item of $(\Omega,\Omega_M)$\footnote{We do not use the term \emph{individual object} in this section to avoid confusing an item of $\Omega$ with an item of the composite system $(\Omega,\Omega_M)$.} be in a pure state represented by the unit vector $|\Psi_{0}\rangle=|\psi\rangle|a_{0}^{M}\rangle \in {\mathscr H}\otimes {\mathscr H}^{M}$, with $|\psi\rangle=\sum_{n=1}^{N}\sum_{\mu=1}^{g_n}c_{n}^{\mu}|a_{n}^{\mu}\rangle$ the unit vector of ${\mathscr H}$ representing a pure state $P$ of $\Omega$. Then an idealized measurement of $A_0$ maps $|\Psi_{0}\rangle$ into a unit vector $|\Psi\rangle$, as follows.
\begin{eqnarray} 
|\Psi_{0}\rangle=|\psi\rangle| a_{0}^{M}\rangle=\sum_{n=1}^{W} \sum_{\mu=1}^{g_{n}} c_{n}^{\mu} |a_{n}^{\mu}\rangle |a_0^{M}\rangle \nonumber \\
\longrightarrow |\Psi\rangle=\sum_{n=1}^{W} \sum_{\mu=1}^{g_{n}}  \alpha_{P, n} c_{n}^{\mu} |a_{n}^{\mu} \rangle |a_{n}^{M}\rangle+\beta_{P, 0} |\psi_{F_0}\rangle |a_0^{M}\rangle . \label{fop2011}
\end{eqnarray}
The coefficients $\alpha_{P, n}$ and $\beta_{P, 0}$ in Eq. (\ref{fop2011}) are given by
\begin{equation} \label{phase_factors}
\left \{
\begin{array}{l}
\alpha_{P, n}=\sqrt{p_{P, A_{0}}^{d}(a_n)}e^{i \theta_{P, n}} \\
\beta_{P, 0} =\sqrt{p^{t}(P,F_0)}e^{i \varphi_{P, 0}}
\end{array}
\right. ,
\end{equation}
where $F_0$ is the physical property $(A_0, \{ a_0\})$, $\theta_{P, n}$ and $\varphi_{P, 0}$ are arbitrary real numbers, and the following equation holds
\begin{equation} 
\sum_{n=1}^{W}\sum_{\mu=1}^{g_n} |\alpha_{P, n}c_{n}^{\mu}|^{2}+|\beta_{P, 0}|^{2}=\langle\Psi|\Psi\rangle=1 .
\end{equation}

It is apparent that assumption AXM modifies the standard description of the measurement process in QM \cite{vn32} by introducing the vector $|\psi_{F_0}\rangle$ which represents the final state of the item of $\Omega$ that is measured whenever the $a_0$ outcome is obtained.

By using Eq. (\ref{fop2011}) one can write the density operator $\rho_{\Psi}=|\Psi\rangle\langle\Psi|$ representing the state of an item of $(\Omega, \Omega^M)$ after the measurement. Hence, one can obtain the density operator  $\tilde{\rho}=Tr_{\Omega^M}\rho_{\Psi}$ representing the state (improper mixture) of an item of $\Omega$ after the measurement (Sect. \ref{esrsub4}). It follows from Eq. (\ref{fop2011}) that
\begin{eqnarray}\label{premeasurement}
\tilde{\rho}=\sum_{n=1}^{W} \langle a_{n}^{M} |\Psi\rangle\langle\Psi| a_{n}^{M} \rangle \nonumber \\
=|\beta_{P, 0}|^{2} |\psi_{F_0}\rangle\langle \psi_{F_0}|+\sum_{n=1}^{W}|\alpha_{P, n}|^{2}\sum_{\mu,\nu=1}^{g_{n}}c_{n}^{\mu}(c_{n}^{\nu})^{*} |a_{n}^{\mu} \rangle\langle a_{n}^{\nu}| \nonumber \\
=|\beta_{P,0}|^{2}| \psi_{F_0} \rangle\langle \psi_{F_0}|+\sum_{n=1}^{W}|\alpha_{P, n}|^{2} P_{n}^{\widehat{A}} \rho_{P} P_{n}^{\widehat{A}} %\nonumber \\
%=p_{S}^{t}(F_0) \rho_{\psi_{F_0}}+\sum_{n=1}^{N} p_{S}^{d}(A_0, \{ a_n \}) P_{n}^{\widehat{A}} \rho_{\psi} P_{n}^{\widehat{A}} 
. \label{partial_GPP}
\end{eqnarray}
It is then natural to generalize Eq. (\ref{partial_GPP}) to every state $S \in {\mathscr P}\cup{\mathscr N}$ of $\Omega$, as follows.
\begin{equation} \label{measurement}
\tilde{\rho}=p^{t}(P,F_0) \rho_{S_{F_0}}+\sum_{n=1}^{W} p_{S,A_{0}}^{d}(a_n) P_{n}^{\widehat{A}} \rho_{S} P_{n}^{\widehat{A}} ,
\end{equation} 
where $\rho_{S_{F_0}}$ is given by Eq. (\ref{gpp_improper}), with $S$ in place of $P$. Equation (\ref{measurement}) thus provides the basic equation for the state transformation induced by a measurement in the ESR model.

Finally, we note that Eq. (\ref{partial_GPP}) modifies the perspective in \cite{gs10}: indeed, we do not obtain a proper mixture as a final state after a nonselective measurement of a generalized observable on an item of $\Omega$ in a pure state $P$ but, rather, an improper mixture.

\subsection{Linear unitary evolution\label{subevo2}}
The evolution of the composite physical system $(\Omega,\Omega^M)$ postulated by assumption AXM depends on the unknown values of the detection probability via the parameters $\alpha_{P, n}$ and $\beta_{P, 0}$ that occur in Eq. (\ref{fop2011}). We may then wonder whether these parameters can be determined in such a way that the evolution is induced by a linear unitary operator. To answer this question let us refer to the symbols introduced in Sect. \ref{subevo1}, consider again a pure state of $(\Omega,\Omega^M)$ represented by the unit vector $|\Psi_0\rangle=|\psi\rangle |a_0^M\rangle$, and denote by $S_{n}^{\mu}$ the pure state of $\Omega$ represented by the unit vector $|a_{n}^{\mu}\rangle$. Then, Eq. (\ref{fop2011}) yields
\begin{equation} \label{fop2004}
|a_{n}^{\mu}\rangle |a_{0}^{M}\rangle \longrightarrow \alpha_{S_{n}^{\mu}, n} |a_{n}^{\mu}\rangle|a_{n}^{M}\rangle+ \beta_{S_{n}^{\mu},0} |(a_{n}^{\mu})_{F_0} \rangle |a_{0}^{M}\rangle .
\end{equation}
Let us assume that $\alpha_{S_{n}^{\mu},n}$ and $\beta_{S_{n}^{\mu},0}$ are real and do not depend on $\mu$, consistently with their physical interpretation as square roots of probabilities, up to a phase factor (Eq. (\ref{phase_factors})). Then, we briefly put $\alpha_{S_{n}^{\mu},n}=\sqrt{p_{S_{n}^{\mu},A_{0}}^{d}(a_n)}=\alpha_n$, $\beta_{S_{n}^{\mu},0}=\sqrt{p^{t}(S_n^{\mu},F_0)}=\beta_{n,0}$ and $|(a_{n}^{\mu})_{F_0} \rangle= |a_{n,0}^{\mu} \rangle$ (hence, %we can rewrite Eq. (\ref{fop2004}) as follows.
%\begin{equation} \label{fop2004_simplified}
%|a_{n}^{\mu}\rangle |a_{0}^{M}\rangle \longrightarrow \alpha_n |a_{n}^{\mu}\rangle |a_{n}^{M}\rangle +\beta_{n0} |a_{n0}^{\mu} \rangle |a_{n}^{M}\rangle,
%\end{equation}
%with 
$|\alpha_n|^{2}+|\beta_{n,0}|^{2}=1$). If we now assume that the evolution of $(\Omega, \Omega^M)$ is induced by a linear unitary operator $U$, we obtain from Eq. (\ref{fop2004})
\begin{displaymath} 
|\Psi_{0}\rangle=|\psi\rangle| a_{0}^{M}\rangle=\sum_{n=1}^{W} \sum_{\mu=1}^{g_{n}} c_{n}^{\mu} |a_{n}^{\mu}\rangle |a_0^{M}\rangle
\end{displaymath}
\begin{displaymath}
\begin{CD}
{} @>>U>|\Psi\rangle= \sum_{n=1}^{W} \sum_{\mu=1}^{g_{n}}c_{n}^{\mu} (\alpha_{n} |a_{n}^{\mu} \rangle |a_{n}^{M}\rangle+\beta_{n, 0} |a_{n,0}^{\mu}\rangle |a_0^{M}\rangle)
\end{CD}
\end{displaymath}
\begin{equation}
=\sum_{n=1}^{W} \sum_{\mu=1}^{g_{n}}\alpha_{n} c_{n}^{\mu}  |a_{n}^{\mu} \rangle |a_{n}^{M}\rangle + \sum_{n=1}^{W} \sum_{\mu=1}^{g_{n}}\beta_{n, 0} c_{n}^{\mu} |a_{n,0}^{\mu}\rangle |a_0^{M}\rangle . \label{fop2011_linearized}
\end{equation}
By comparing Eq. (\ref{fop2011_linearized}) with Eq. (\ref{fop2011}) we conclude that the general evolution described by Eq. (\ref{fop2011}) is linear whenever the following conditions hold for every $|\psi\rangle\in {\mathscr H}$
\begin{equation} \label{conditions}
\left \{
\begin{array}{l}
\alpha_{P, n}=\alpha_{n} \\
\beta_{P, 0} |\psi_{F_0} \rangle =\sum_{n=1}^{W} \sum_{\mu=1}^{g_{n}} \beta_{n, 0} c_{n}^{\mu} |a_{n,0}^{\mu}\rangle 
\end{array}
\right..
\end{equation}
We can thus maintain that  these conditions are always satisfied and that $(\Omega, \Omega^M)$ undergoes linear unitary evolution.\footnote{The evolution described by Eq. (\ref{fop2011_linearized}) coincides with the evolution postulated in \cite{gp04}. Hence the latter is a special case of the general evolution described by Eq. (\ref{fop2011}).} The density operator $\tilde{\rho}$ is given in this case by 
%. We get at once from Eqs. (\ref{partial_GPP}), (\ref{fop2011_linearized}) and (\ref{conditions})
\begin{equation} \label{simple}
\tilde{\rho}=|\beta_{P, 0}|^{2} |\psi_{F_{0}}\rangle\langle\psi_{F_{0}}|+\sum_{n=1}^{W}|\alpha_{n}|^{2} P_{n}^{\widehat{A}} \rho_{P} P_{n}^{\widehat{A}} , 
\end{equation}
where
\begin{equation}
|\beta_{P, 0}|=|\sum_{n=1}^{W} \sum_{\mu=1}^{g_{n}} \beta_{n, 0}c_{n}^{\mu}  |a_{n,0}^{\mu}\rangle |
\end{equation}
and 
\begin{equation} 
|\psi_{F_0} \rangle=\frac{\sum_{n=1}^{W} \sum_{\mu=1}^{g_{n}} \beta_{n, 0}c_{n}^{\mu} |a_{n,0}^{\mu}\rangle}{|\beta_{P, 0}|} .
\end{equation}
One may now wonder whether the linear unitary evolution of $(\Omega, \Omega^M)$ could lead to linear unitary evolution of the subsystem $\Omega$ in the ESR model. Eq. (\ref{simple}) implies that this is not the case and that the reduced dynamics induced by a measurement is necessarily nonlinear. Let us prove this statement in the case of a nondegenerate observable (the generalization is immediate). By putting
%$\tilde{\rho}$ in Eq. (\ref{simple}) can be reduced to a one--dimensional projection operator, thus representing a pure state (which is a necessary condition for linear unitary evolution of $|\psi\rangle$ in the course of the measurement). For the sake of simplicity, let us consider the special case of a nondegenerate observable, put
 $p_{n}=|\alpha_n c_n|^{2}$, $p=|\beta_{P, 0}|^{2}$ and $|\psi_{F_{0}}\rangle=\sum_{n=1}^{W} k_{n}|a_n \rangle$, we get from Eq. (\ref{simple})
\begin{eqnarray}
\tilde{\rho}=p\sum_{n,n'=1}^{W} k_{n}k_{n'}^{*} |a_n\rangle\langle a_{n'}|+\sum_{n=1}^{W}p_{n}|a_n\rangle\langle a_{n}| \nonumber \\
=\sum_{n,n'=1,n\ne n'}^{W} p k_{n}k_{n'}^{*} |a_n\rangle\langle a_{n'}|+\sum_{n=1}^{W} (p_{n}+p|k_{n}|^{2}) |a_n\rangle\langle a_{n}| \label{nogo}
\end{eqnarray}
%Eq. (\ref{nogo}) must be compared with the general expression of 
If linearity holds, $\tilde{\rho}$ represents a pure state: hence it reduces to
a one--dimensional projection operator $Q$ on $\mathscr H$. Let us put $Q=|\phi\rangle\langle\phi|$ and $|\phi\rangle=\sum_{n=1}^{W}d_n |a_n\rangle$. Then we get
\begin{equation}
Q=\sum_{n,n'=1,n\ne n'}^{W} d_n d_{n'}^{*} |a_n\rangle\langle a_{n'}|+\sum_{n=1}^{W} d_n d_{n}^{*} |a_n\rangle\langle a_{n}| ,
\end{equation}
%Hence, $\tilde{\rho}$ represents a pure state iff 
that is, coefficients $d_1, d_2,\ldots,d_{W}$ exist such that, for every $n,n'=1,2,\ldots,W$, $n \ne n'$,
\begin{equation} \label{nopure1}
d_n d_{n'}^{*}=p k_{n}k_{n'}^{*} ,
\end{equation}
while, for every $n=1,2,\ldots,W$,
\begin{equation} \label{nopure2}
d_n d_{n}^{*}=p_{n}+p|k_{n}|^{2} .
\end{equation}
Eqs. (\ref{nopure1}) and (\ref{nopure2}) imply that, for every $n,n'=1,2,\ldots,W, n \ne n'$, the following condition holds. 
%\begin{equation}
%\sqrt{p_{n}+p|k_{n}(\psi)|^{2}}\sqrt{p_{n'}+p|k_{n'}(\psi)|^{2}}= p |k_{n}(\psi)k_{n'}^{*}(\psi)| ,
%\end{equation}
%that is,
\begin{equation} \label{bo}
(p_{n}+p|k_{n}|^{2})(p_{n'}+p|k_{n'}|^{2})= p^2 |k_{n}|^{2} |k_{n'}|^{2} .
\end{equation}
Eq. (\ref{bo}) is satisfied iff, for every $n=1,2,\ldots,W$, $d_n=\sqrt{p_{S_{n}^{\mu},A_{0}}(a_n)}=0$, hence $p_n=0$, which implies that no detection occurs. We conclude that the time evolution induced by a measurement procedure on an item of $\Omega$ is necessarily nonlinear, as stated.

\subsection{General assumptions on time evolution\label{subevo3}}
Our treatment in Sects. \ref{subevo1} and \ref{subevo2} allows us to draw, if linear Hamiltonian evolution of the composite system $(\Omega, \Omega^{M})$ is postulated, the following commutative diagram
\begin{equation} \label{commutative}
\begin{CD}
\rho_{\Psi_0}=|\psi\rangle\langle\psi|\otimes |a_{0}^{M}\rangle\langle a_{0}^{M}| @>>U> \rho_{\Psi}=|\Psi\rangle\langle\Psi| \\
@V{Tr_{\Omega^M}}VV                    @VV{Tr_{\Omega^{M}}}V \\
\rho_{P}=|\psi\rangle\langle\psi| @>>> \tilde{\rho}  
\end{CD}
\end{equation}
The density operator $\tilde{\rho}$ is not a projection operator (Sect. \ref{subevo2}). To be precise, $\tilde{\rho}$ is diagonal in a basis $\{ |b_m \rangle\}_{m}$ %different from $\{ |a_n^{\mu} \rangle\}_{n, \mu}$, 
in which it takes the form $\tilde{\rho}=\sum_{m} p_{m}|b_m \rangle \langle b_m |$, where at least two values $r$ and $s$ exist such that $p_{r} \ne 0 \ne p_{s}$. Hence the pure state represented by $\rho_{P}$ evolves into the improper mixture represented by %(or \emph{generalized pure state}) 
$\tilde{\rho}$, and its evolution is not linear, even if the evolution of the composite system $(\Omega,\Omega_{M})$ is linear.

To avoid contradiction preserving linear evolution as far as possible in the ESR model, one can assume the following general rules for the time evolution of a state $S \in {\mathscr P}\cup {\mathscr N}$ represented by the density operator $\rho_{S}(t)$ at time $t$.

(i) {\it Closed systems:} linear evolution ruled by the von Neumann--Liouville equation
\begin{equation}
i \hbar \frac{d\rho_{S}(t)}{dt}=[\hat{H},\rho_{S}(t)] ,
\end{equation}
where $\widehat{H}$ is a self--adjoint Hamiltonian.

(ii) {\it Open systems:} non--necessarily linear evolution, as exemplified by the mapping $\rho_{P} \longrightarrow\tilde{\rho}$ in the diagram above. Whenever the open system can be considered as a subsystem of a closed system, its dynamics can be deduced from the dynamics of the closed system, and it may be linear or not depending on the Hamiltonian of the closed system.

One thus converges to a standard perspective in QM \cite{d76,bp02}, but avoids the problems that occur in QM because of nonobjectivity, as we show in the next section.

It remains to discuss time evolution in the case of proper mixtures. Let us therefore consider a \emph{generalized proper mixture} $M$ of the states $S_1, S_2, \ldots \in {\mathscr P}\cup {\mathscr N}$, with probabilities $p_1,p_2,\ldots$, respectively. In this case, it is natural to assume that $S_1,S_2,\ldots$ change with time according to the rules supplied above, while $p_1,p_2,\ldots$ do not change. This assumption is sufficient to provide the desired evolution. Moreover, it implies that the state transformation induced on a proper mixture by an idealized measurement of a discrete generalized observable $A_0$ can be deduced from Eq. (\ref{measurement}). The explicit form of this transformation has been studied in some previous papers \cite{gs12,gs10}, and we do not report it here for the sake of brevity.

\subsection{Justifying the generalized L\"{u}ders postulate\label{subevo4}}
The change of state of $(\Omega ,\Omega ^{M})$ postulated by Eq. (\ref{premeasurement}) has been hypothesized bearing in mind the final state of $\Omega $ specified by Eq. (\ref{gpp_improper}). Conversely, let us show that Eq. (\ref{premeasurement}) justifies the special form of GLP specified by Eq. (\ref{gpp_improper}) if one resorts to the interpretation of the measurement process in terms of hidden variables provided in Sect. \ref{submicro1}.

The final state after an idealized measurement of a discrete generalized observable $A_{0}$ on an item $\alpha $ of $\Omega $ in a pure state $P$ is represented by the density operator $\tilde{\rho}$ in Eq. (\ref{premeasurement}). This operator can be written in the form 
\begin{equation}
\tilde{\rho}=|\beta _{P,0}|^{2}|\psi _{F_{0}}\rangle \langle \psi_{F_{0}}|+\sum_{n=1}^{W}\gamma_{P,n}\frac{P_{n}^{\widehat{A}}\rho _{P}P_{n}^{\widehat{A}}}{Tr[P_{n}^{\widehat{A}}\rho _{P}P_{n}^{\widehat{A}}]} \label{justif}
\end{equation}
where $\gamma_{P,n}=|\alpha _{P,n}|^{2}Tr[P_{n}^{\widehat{A}}\rho _{P}P_{n}^{\widehat{A}}]$. Equation (\ref{justif}) provides a decomposition of $\tilde{\rho}$ in terms of pure states. But the coefficients $\gamma_{P,1}$, $\gamma_{P,2}$, \ldots, $\gamma_{P,W}$ that occur in it do not represent probabilities of the corresponding pure states. Rather, they represent the overall probabilities that the physical properties $F_{1}$, $F_{2}$, \ldots, $F_{W}$, respectively, be displayed in the measurement. These probabilities are epistemic in the ESR model whenever this model is objective (Sect. \ref{submicro2}). In this case they formalize our \emph{a priori} ignorance of the outcome that will be obtained in the measurement. Whenever the display of
the apparatus measuring $A_{0}$ is observed, this ignorance is reduced, and we are informed that $\alpha $ displays a specific property, say $F_{k}=(A_{0},\{a_{k}\})$. Thus we can update our information about the properties of $\alpha $. Let $k\neq 0$. Then, according to the hidden variables theory in Sect. \ref{submicro1}, $\alpha $ possesses the microscopic
property $f_{k}=\varphi ^{-1}(F_{k})$. Therefore, if the measurement is repeated and $\alpha$ is detected, it must yield the same result with certainty. This means that the conditional on detection probability of $F_{k}$ after the first measurement is 1, which is just what occurs if $\alpha $ is in the pure state represented by $\frac{P_{k}^{\widehat{A}}\rho _{P}P_{k}^{%
\widehat{A}}}{Tr[P_{k}^{\widehat{A}}\rho _{P}P_{k}^{\widehat{A}}]}$ after the measurement, as predicted by Eq. (\ref{gpp_improper}). Yet, no ``collapse of the wave function'' occurs, because $F_{k}$ is not ``brought into existence'' by the measurement, as in QM. Let $k=0$. Then, we deduce that the set of microscopic properties possessed \ by $\alpha $ is such that no-detection
may occur. If the measurement is repeated, either no-detection occurs again, or one of the physical properties $F_{1}$, $F_{2}$, \ldots\, $F_{W}$ is displayed. If we assume that the conditional on detection probability of $F_{k}$ is given by $Tr[|\psi _{F_{0}}\rangle \langle \psi _{F_{0}}|P_{k}^{\widehat{A}}]$, we can maintain that the state of $\alpha $ after the measurement is the pure
state represented by $|\psi _{F_{0}}\rangle $, as predicted by Eq. (\ref{gpp_improper}).

To close, let us observe that the above reasoning does not justify the general form of GLP provided by Eq. (\ref{genpost_dis_W}). This
justification would require a generalization of our arguments. Indeed, consider again a discrete generalized observable $A_{0}$ and an item of $\Omega$ in the pure state $P$. Performing an idealized measurement of $A_{0} $ which yields one of the outcomes $v(a_0)$, $v(a_1)$, $v(a_2)$, \ldots, $v(a_W)$ is equivalent to measuring all physical properties in the set ${\mathscr F}_{A_{0}}=\{(A_{0},\Sigma )\ |\ \Sigma \subset \Xi _{0}\}$ associated with $A_{0}$. Each of these measurements yields as final state one of the states represented by the density operators in Eq. (\ref{gpp_improper}). Hence the state after one of them coincides with the state predicted
by GLP only for physical properties in the subset $\{(A_{0},\{a_{0}\})\ | \ a_{n}\in \Xi _{0}\}\subset {\mathscr F}_{A_{0}}$. In different words, the
measurement of the physical property $F=(A_{0},\Sigma )\in {\mathscr F}_{A_{0}}$ following from an idealized measurement of $A_0$ is an idealized measurement in the sense established by Eq. (\ref{genpost_dis_W}) only if $\Sigma =\{a_{n}\}$, with $n=0,1,2,\ldots ,W$. To justify the general form of GLP one should consider idealized measurements of generalized observables obtained from $A_{0}$ by grouping together sets
of eigenvalues of $A_{0}$ and considering each set as a single eigenvalue. This procedure is rather obvious and we do not discuss it here for the sake of brevity.

\section{Conclusions: advantages and limits of the ESR model\label{conclusions}}
As we have seen in the previous sections, every physical system $\Omega$ is associated, according to the ESR model, not only with a Hilbert space $\mathscr H$, but also with a detection probability $p^{d}(S,F)$ depending on the state $S$ of the individual object $\alpha$ that is considered and on the physical property $F$ of $\alpha$ that is measured. The main limit of this description is that we have as yet no theory for the detection probability, even if its introduction can be justified by a hidden variables theory for the measurement process (Sect. \ref{hv}). Hence the basic assumptions of the ESR model (Sect. \ref{esrsub3}) must be considered as {\it a priori} hypotheses, to be accepted or not because of their physical consequences, explanatory power and predictions. Moreover, the values of the detection probability cannot be predicted, hence they occur in the ESR model as parameters to be determined experimentally for every physical system $\Omega$, state $S$ and property $F$. We intuitively expect very different values for different physical systems: {\it e.g.}, very cloe to 1 in the case of massive physical systems, as heavy ions, sensibly different from 1 in the case of lighter particles, as electrons or photons. 

Whenever the ESR model is objective in the sense specified in Sect. \ref{objectivity}, it exhibits some physically interesting features. Let us summarize some of them.

(i) The objectification problem and the paradoxes ensuing from it are avoided because of objectivity. An epistemic interpretation of quantum probabilities becomes possible, and no ambiguity occurs in the interpretation of mixtures because proper and improper mixtures have different mathematical representations. Furthermore, the ESR model supports a reinterpretation of standard quantum logic which makes it compatible with classical logic \cite{gserke}. The price of these achievements is a mathematical description of physical entities that is more complicated than the mathematical description provided by QM, but the latter is recovered within the ESR model (Sect. \ref{esrsub4}).

(ii) The relationship of the predictions of the ESR model with the predictions of QM is not trivial. In the case of experiments on the CHSH inequality, the experimenters must check the four terms of the sum that occurs in this inequality, each representing the expectation value of the product of two compatible dichotomic observables in a given state $S$, on four sets ${\mathscr U}_1$, ${\mathscr U}_2$, ${\mathscr U}_3$, ${\mathscr U}_4$ of individual objects in the state $S$. But the low efficiency of real detectors obliges them to take into account, rather than these sets, the subsets ${\mathscr U}_1^{d}\subset {\mathscr U}_1$, ${\mathscr U}_2^{d}\subset {\mathscr U}_2$, ${\mathscr U}_3^{d}\subset {\mathscr U}_3$, ${\mathscr U}_4^{d}\subset {\mathscr U}_4$ of the individual objects that are detected. The CHSH inequality is then checked (and found to be violated, consistently with the predictions of QM) by assuming that, for every $i \in \{ 1,2,3,4\}$, ${\mathscr U}_i^{d}$ is a fair sample of ${\mathscr U}_i$ ({\it fair sampling}, or {\it no-enhancement}, assumption \cite{s04,s05}). According to the ESR model, instead, one must consider, for every $i \in \{ 1,2,3,4\}$, also the subset ${\mathscr U}_i^{d_0}\subset {\mathscr U}_{i}$ of all individual objects that would be detected if idealized measurements were performed, which is such that ${\mathscr U}_i^{d}\subset {\mathscr U}_i^{d_0}\subset {\mathscr U}_{i}$. Generally, ${\mathscr U}_i^{d_0}$ is not a fair sample of ${\mathscr U}_i$, but the statistical predictions of the ESR model on ${\mathscr U}_i^{d_0}$ are identical to the statistical predictions of QM on ${\mathscr U}_i$. If one then considers real measurements and assume that ${\mathscr U}_i^{d}$ is a fair sample of ${\mathscr U}_i^{d_0}$, one expects from Aspect's experiments exactly the same results that are expected when the quantum description of the experiment is adopted: hence, a violation of the CHSH inequality. This predictions does not depend on the objectivity of the ESR model. But if this model can supply an objective description of the physical situation ({\it i.e.}, if the efficiency limits reported in Sect. \ref{submicro2} are respected), then the results of the experiments can be explained by assuming that  ${\mathscr U}_i^{d_0}$, hence ${\mathscr U}_i^{d}$, is not a fair sample of ${\mathscr U}_i$. One thus obtains an explanation in which nonlocality plays no role \cite{g07,s13}.

An analogous but simpler situation occurs in the case of the various versions of the two-slits experiment. Indeed, also in this class of experiments only detected individual objects are taken into account. Hence, the ESR model predicts interference, exactly as in QM, whenever idealized measurements are performed (of course, the interference fringes could be modified in the case of real measurements). This result is not counterintuitive, because the ESR model does not imply any picture of individual objects as point-like classical or semi-classical particles with trajectories (Sect. \ref{hv}). Moreover, the prediction that there will be interference does not depend on the values of the detection probability, hence on the objectivity of the ESR model.
  
One can contrive, however, experiments that take into account all individual objects that are produced. The predictions of the ESR model are then generally different from the predictions of QM, but can be very close to them if all the values of the detection probability that occur in the experiment are close to 1. The predictions of the two theories, however,  may substantially differ in specific cases: for instance, when proper mixtures are considered \cite{gs12}. In this case one can devise experiments to check which description is correct.

(iii) The transformation of state induced by an idealized measurement is described in the ESR model by a postulate that generalizes the L\"{u}ders postulate of QM (Sect. \ref{esrsub5}). If the ESR model is objective, this postulate does not imply any actualization of physical properties following from a ``collapse of the wave function''. Rather, it provides the final state of an individual object after an interaction with an idealized macroscopic measuring device. Hence this description can be used as a starting point for hypothesizing the general laws of time evolution in the ESR model, which then basically reproduces standard quantum laws (Sect. \ref{subevo3}).

To close let us add some remarks on the literature.

First of all, we stress that the basic ideas of the ESR model make it very different from the models aiming to exploit the inefficiency of the detectors, or other loopholes, to explain the results obtained by the experiments on Bell's inequalities in a perspective of ``local realism'' which is alternative to QM. Indeed, the ESR model does not reject QM but recovers its mathematical formalism reinterpreting quantum probability, as explained in Sect.\ref{intro}. There are instead strong similarities between the ESR model and some hidden variables models which assume that the no-detection outcome is a possible result of an exact measuring process, and that the efficiency of the detectors used in the experiments depends on the hidden variables \cite{sf02,f82a,f82b,f89,f94}. These assumptions indeed are consistent with the introduction of no--registration outcomes in the ESR model and with the hidden variables theory in Sect. \ref{hv}. Moreover, a reinterpretation of quantum probabilities as conditional on detection is implicitly introduced (but not explicitly stated) when local models for the GHZ experiment are constructed \cite{sf02}, which matches AX 3. However, these local models (dubbed by Fine ``prism models'') mainly aim to show that the experimental results obtained in Aspect's and similar experiments can be explained avoiding nonlocality, and do not provide a general theory vindicating locality. The ESR model supplies instead such a theory, constructing a framework in which ``prism'' and similar models can be placed.

Secondly, we recall that there have been many scholars who attempted to invalidate the theoretical and/or the experimental proofs of nonlocality of QM. In particular, the proofs of some ``no--go'' theorems were questioned by several authors in the framework of a statistical interpretation of QM that does not refer explicitly to individual objects
\cite{a05,k09,k05,h03,h05,k12,k12a,k12b,a09,knew1,knew2}. For instance, Khrennikov introduced a pre-quantum model of the wave type ({\it pre-quantum classical statistical field theory}, or PCSFT) which avoids nonlocality but recovers contextuality when combined with detection by detectors with a threshold ({\it threshold signal detection model}, or TSD). According to these combined models contextuality is a consequence of the limited efficiency of the detectors. More recently Khrennikov, accepting von Neumann's theory of measurement, suggested that the quantum probabilities used in the proofs of the CHSH inequalities and related experiments should be interpreted as conditional (quantum) probabilities, where conditioning occurs with respect to fixed experimental settings \cite{kprep01}. On this basis, he criticized the widespread belief that the Bell--type experiments disprove local realism \cite{kprep02} and upheld that all quantum probabilities can be modeled as classical conditional probabilities \cite{kprep03}.

The view of the scholars mentioned above, however, is different from the perspective of the ESR model. Indeed, also this model reinterprets quantum probabilities as conditional (on detection). Yet, conditioning does not occur with respect to fixed {\it real} experimental settings but with respect to {\it idealized} measurements which test properties of individual objects. In this kind of measurements there are not thresholds or external sources of randomness, and no--detection can be ascribed to the set of microscopic properties possessed by the individual object that is considered according to the hidden variables theory of measurement expounded in Sect. \ref{hv}. Hence, not only nonlocality, but also contextuality is avoided (if suitable conditions are fulfilled, see Sect. \ref{submicro2}). These features follow from the fundamental choice of accepting a ``realistic'' interpretation of QM (see footnote 1), which maintains that QM deals with individual objects and their properties. Indeed this choice raises the objectification problem that does not occur in a purely statistical interpretation of QM, as we have already observed in Sect. \ref{intro}, and the assumptions of the ESR model (Sect. \ref{esrsub3}) mainly aim to solve this problem. 

Notwithstanding the differences pointed out above, there are some interesting similarities between the ESR model and the foregoing approaches. In fact, (local) contextuality similar to Khrennikov's would occur also in the ESR model if real detectors with thresholds were considered instead of idealized measurement devices \cite{gs08,gs10}. Moreover, an epistemic interpretation of probabilities, hence a reinterpretation of quantum probability as classical conditional probability, is possible if the ESR model is objective (Sect. \ref{submicro2}). Thus, Khrennikov's proposals share several important features with the ESR model.

%Secondly, we recall that there have been many scholars who accepted QM but attempted to invalidate the theoretical and/or the experimental proofs of nonlocality. In particular the standard proofs of some ``no--go'' theorems were questioned by several authors in the framework of a statistical interpretation of QM that maintains the orthodox interpretation of quantum probabilities but avoids referring explicitly  to individual objects \cite{a05,k09,k05,h03,h05,k12,k12a,k12b,a09,knew1,knew2}. The view of these scholars therefore seems very different from the perspective of the ESR model. However, this is not necessarily true. For instance, Khrennikov introduced a pre-quantum model of the wave type ({\it pre-quantum classical statistical field theory}, or PCSFT) which avoids nonlocality but recovers contextuality when combined with detection by detectors with a threshold ({\it threshold signal detection model}, or TSD). According to these combined models contextuality is a consequence of the limited efficiency of the detectors. But such a contextuality could occur also in the ESR model if real detectors with thresholds were considered instead of idealized measurement devices \cite{gs08,gs10}. Moreover, a reinterpretation of quantum probability as conditional on detection seems implicit in PCSFT plus TSD \cite{gps13}. Thus, Khrennikov's models share several important features with the ESR model.
 
\vspace{.5cm}
\noindent 
\section*{Acknowledgements}
This work was supported by the Natural Science Foundations of China (11171301 and 10771191) and by the Doctoral Programs Foundation of Ministry of Education of China (J20130061).


\begin{thebibliography}{99}

\bibitem{ba70} Ballentine, L.E.: The statistical interpretation of quantum mechanics. Rev. Mod. Phys. \textbf{42}, 358--381 (1970).

\bibitem{blm96} Busch, P., Lahti, P.J., Mittelstaedt, P.: The Quantum Theory of Measurement. Springer, Berlin (1996).

%\bibitem{bgl96} Busch, P., Grabowski, M., Lahti, P.J.: Operational Quantum Physics. Springer, Berlin (1996).

\bibitem{l83} Ludwig, G.: Foundations of Quantum Mechanics I. Springer, Berlin (1983).

\bibitem{me93} Mermin, N.D.: Hidden variables and the two theorems of John Bell. Rev. Mod. Phys. \textbf{65}, 803--815 (1993).

\bibitem{b66} Bell, J.S.: On the problem of hidden variables in quantum mechanics. Rev. Mod. Phys. \textbf{38}, 447--452 (1966).

\bibitem{ks67} Kochen, S., Specker, E.P.: The problem of hidden variables in quantum mechanics. J. Math. Mech. \textbf{17}, 59--87 (1967).

\bibitem{b64} Bell, J.S.: On the Einstein-Podolsky-Rosen paradox. Physics \textbf{1}, 195--200 (1964).

\bibitem{aetal82} Aspect, A., Grangier, P., Roger, G.: Experimental realization of Einstein-Podolsky-Rosen-Bohm Gedankenexperiment: A new violation of Bell's inequalities. Phys. Rev. Lett. \textbf{49}, 91--94 (1982).

\bibitem{a82a} Aspect, A., Dalibard, J., Roger, G.: Experimental test of Bell's inequalities using time-varying analyzers. Phys. Rev. Lett. \textbf{49}, 1804--1807 (1982).

\bibitem{ge05} Genovese, M.: Research on hidden variables theories: A review of recent progresses. Phys. Repts. \textbf{413}, 319--396 (2005).

\bibitem{desp76} d'Espagnat, B.: Conceptual Foundations of Quantum Mechanics. Addison Wesley, New York (1976).

\bibitem{tb05} Timpson C.G., Brown H.R.: Proper and improper separability. Int. J. Quant. Inf. \textbf{3}, 679--690 (2005).

\bibitem{gs12} Garola, C., Sozzo, S.: Extended representations of observables and states for a noncontextual reinterpretation of QM. J. Phys. A: Math. Theor. \textbf{45}, 075303 (2012).

\bibitem{dcgg04} Dalla Chiara, M. L., Giuntini, R., Greechie, R.: Reasoning in quantum theory. Kluwer, Dordrecht (2004).

\bibitem{b52} Bohm, D.: A suggested interpretation of quantum theory in terms of ``hidden variables''. Phys. Rev. \textbf{85}, 166--179 (1952).



\bibitem{z99} Zeilinger, A.:  A foundational principle for quantum mechanics. Found. Phys. \textbf{29}, 631--643 (1999).

\bibitem{cbh03} Clifton, R., Bub, J., Halvorson, H.: Characterizing quantum theory in terms of information theoretic constraints. Found. Phys. \textbf{33}, 1561 (2003).

\bibitem{cfs02a} Caves, C.M., Fuchs, C.A., Schack, R.: Conditions for compatibility of quantum state assignments. Phys. Rev. A \textbf{66}, 062111 pp. 1--11 (2002).

\bibitem{cfs02b} Caves, C.M., Fuchs, C.A., Schack, R.: Unknown quantum states: the quantum de Finetti representation. J. Math. Phys. \textbf{43} 4537 (2002).

\bibitem{fs04} Fuchs, C.A., Schack, R.: Unknown quantum states and operations, a Bayesian view. Lect. Not. Phys. \textbf{649}, 147--187 (2004).

%\bibitem{d10a} D'Ariano, G.M.: Probabilistic theories: what is special about quantum mechanics?. In: Philosophy of Quantum Information and Entanglement. Bokulich, A., Jaeger, G. (Eds) pp. 85--126, Cambridge University Press, Cambridge, UK (2010).

%\bibitem{d10b} D'Ariano, G.M.: On the ``principle of the quantumness'', the quantumness of Relativity, and the computational grand-unification. In: CP1232 Quantum Theory: Reconsideration of Foundations 5. Khrennikov, A.Yu. (Eds.)  pp. 3, AIP, Melville, New York (2010).  

\bibitem{gp04} Garola, C., Pykacz, J.: Locality and measurements within the SR model for an objective interpretation of quantum mechanics. Found. Phys. \textbf{34}, 449--475 (2004).

\bibitem{gp13} Garola, C., Persano, M.: Embedding quantum mechanics into a broader noncontextual theory. Found. Sci. \textbf{19}, 217--239 (2014).

\bibitem{gs96a} Garola, C., Solombrino, L.: The theoretical apparatus of semantic realism: a new language for classical and quantum physics. Found. Phys. \textbf{26}, 1121--1164 (1996)

\bibitem{gs96b} Garola, C., Solombrino, L.: Semantic realism versus EPR-like paradoxes: the Furry, Bohm-Aharonov, and Bell paradoxes. Found. Phys. \textbf{26}, 1329--1356 (1996).

\bibitem{ga02} Garola, C.: A simple model for an objective interpretation of quantum mechanics. Found. Phys. \textbf{32}, 1597--1615 (2002).

\bibitem{gshum} Garola, C., Sozzo, S.: Realistic aspects in the standard interpretation of quantum mechanics. Humana.mente. J. Phil. Stud. \textbf{13}, 81--101 (2010).

\bibitem{ga03} Garola, C.: Embedding quantum mechanics into an objective framework. Found. Phys. Lett. 16, 605--612 (2003).

\bibitem{g07} Garola, C.: The ESR model: reinterpreting quantum probabilities within a realistic and local framework. In: Adenier, G., {\it et al.} (eds.) Quantum Theory: Reconsideration of Foundations-4, pp. 247--252. American Institute of Physics, Ser. Conference Proceedings 962, Melville (2007)

\bibitem{s07} Sozzo, S.: Modified BCHSH inequalities within the ESR model. In: Adenier, G., {\it et al.} (eds.) Quantum Theory: Reconsideration of Foundations-4, pp. 334--338. American Institute of Physics, Ser. Conference Proceedings 962, Melville (2007)

\bibitem{gs09} Garola, C., Sozzo, S.: The ESR model: a proposal for a noncontextual and local Hilbert space extension of QM. Europhys. Lett. \textbf{86}, 20009 (2009).

\bibitem{gs08} Garola, C., Sozzo, S.: Embedding quantum mechanics into a broader noncontextual theory: a conciliatory result. Int. J. Theor. Phys. \textbf{49}, 3101--3117 (2010).

\bibitem{sg08} Sozzo, S., Garola, C.: A Hilbert space representation of generalized observables and measurement processes in the ESR model. Int. J. Theor. Phys. \textbf{49}, 3262--3270 (2010).

\bibitem{gs10} Garola, C., Sozzo, S.: Generalized observables, Bell's inequalities and mixtures in the ESR model for QM. Found. of Phys. \textbf{41}, 424--449 (2011).

\bibitem{gs10b} Garola, C., Sozzo, S.: The modified Bell inequality and its physical implications in the ESR model. Int. J. Theor. Phys. \textbf{50}, 3787--3799 (2011).

\bibitem{gs10c} Garola, C., Sozzo, S.: Representation and interpretation of quantum mixtures in the ESR model. Theor. Math. Phys. \textbf{168}, 912--923 (2011).

\bibitem{gps13} Garola, C., Persano, M., Pykacz, J., Sozzo, S.: Finite local models for the GHZ experiment. Int. J. Theor. Phys. \textbf{53}, 622--644 (2014).

\bibitem{gserke} Garola, C., Sozzo, S.: Recovering nonstandard logics within an extended classical framework. Erkenntnis \textbf{78}, 399--419 (2013).

\bibitem{s13} Sozzo, S.: The quantum harmonic oscillator in the ESR model. Found. Phys. \textbf{43}, 792--804 (2013).

\bibitem{g15} Garola, C.: A survey of the ESR model for an objective reinterpretation of quantum mechanics. Int. J. Theor. Phys., DOI:10.1007/s10773-015-2618-y.

\bibitem{chsh69} Clauser, J.F., Horne, M.A., Shimony, A., Holt, R.A.: Proposed experiment to test local hidden-variable theories. Phys. Rev. Lett. {\bf 23}, 880--884 (1969).

\bibitem{ghsz82} Greenberger, D.M., Horne, M.A., Shimony, A., Zeilinger, A.: Bell’s theorem without inequalities. Am. J. Phys. {\bf 58}, 1131--1143 (1982).

\bibitem{bc81} Beltrametti, E.G., Cassinelli, G.: The Logic of Quantum Mechanics. Addison--Wesley, Reading, MA (1981).

\bibitem{b53} Braithwaite, R.B.: Scientific Explanation. Cambridge University Press, Cambridge (1953).

\bibitem{h65} Hempel, C.G.: Aspects of Scientific Explanation. Free Press, New York (1965).

%\bibitem{l76} Ludwig, G.: (1976).

%\bibitem{2b57} Bohm, D., Aharonov, Y.: Discussion of experimental proof for the paradox of Einstein, Rosen, and Podolsky. Phys. Rev. \textbf{108}, 1070-–1076 (1957).

%\bibitem{bvn36} Birkhoff, G., von Neumann, J.: The logic of quantum mechanics. Ann. Math. \textbf{37}, 823--843 (1936).

%\bibitem{m63} Mackey, G.W.: The Mathematical Foundations of Quantum Mechanics. W. A. Benjamin, New York, NY (1963).

%\bibitem{j68} Jauch, J.M.: Foundations of Quantum Mechanics. Addison Wesley, Reading, MA (1968).

%\bibitem{v68} Varadarajan, V.: Geometry of Quantum Theory I \& II. von Nostrand, Princeton, NJ (1968).

%\bibitem{p76} Piron, C.: Foundations of Quantum Physics. Benjamin, Reading, MA (1976).

\bibitem{a99} Aerts, D.: Foundations of quantum physics: a general realistic and operational approach. Int. J. Theor. Phys. \textbf{38}, 289--358 (1999).

%\bibitem{jvnw34} Jordan, P., von Neumann, J., Wigner, E.P: On an algebraic generalization of the quantum mechanical formalism. Ann. Math. \textbf{35}, 29--64 (1934).

%\bibitem{s47} Segal, I.E.: Irreducible representations of operator algebras. Bull. Am. Math. Soc. \textbf{53}, 73--88 (1947).


%\bibitem{bs96} Busch, P., Shimony, A.: Insolubility of the quantum measurement problem for unsharp observables. Stud. His. Phil. Mod. Phys. \textbf{27B}, 397--404 (1996).

%\bibitem{b98} Busch, P.: Can `unsharp objectification' solve the quantum measurement problem?. Int. J. Theor. Phys. \textbf{37}, 241--247 (1998).


%\bibitem{r98} R\`edei, M.: Quantum Logic in Algebraic Approach. Kluwer, Dordrecht (1998).




%\bibitem{s04} Santos, E.: The failure to perform a loophole-free test of Bell’s inequality supports local realism. Found. Phys. 34, 1643--1673 (2004).

%\bibitem{ghz89} Greenberger D.M., Horne, M.A., Zeilinger, A.: Going beyond Bell's theorem. In: Bell's Theorem, Quantum Theory and Conceptions of the Universe. Kafatos, M. (Ed) pp. 73--76, Kluwer, Dordrecht (1989).

%\bibitem{ghsz90} Greenberger, D.M., Horne, M.A., Shimony, A., Zeilinger, A.: Bell's theorem without inequalities. Am. J. Phys. \textbf{58} 1131--1143 (1990).

\bibitem{sf02} Szab\'{o} L.E., Fine, A.: A local hidden variable theory for the GHZ experiment. Phys. Lett. A \textbf{295}, 229--240 (2002).

\bibitem{vn32} von Neumann, J.: Mathematical Foundations of Quantum Mechanics. Princeton University Press, Princeton (1932). 



\bibitem{d76} Davies, E.B.: Quantum Theory of Open Systems. Academic Press, London (1976).

\bibitem{bp02} Breuer, H.P., Petruccione, F.: The Theory of Open Quantum Systems. Oxford University Press, Oxford (2002).


\bibitem{s04} Santos, E.: The failure to perform a loophole-free test of Bell’s inequality supports local realism. Found. Phys. {\bf 34}, 1643--1673 (2004).

\bibitem{s05} Santos, E.: Bell’s theorem and the experiments: increasing empirical support for local realism? Stud. Hist. Philos. Mod. Phys. {\bf 36}, 544--565 (2005).

\bibitem{f82a} Fine, A.: Some local models for correlation experiments. Synthese \textbf{50}, 279--294 (1982).

\bibitem{f82b} Fine, A.: Hidden variables, joint probability and the Bell inequalities. Phys. Rev. Lett. \textbf{48}, 291--295 (1982).

\bibitem{f89} Fine, A.: Correlations and efficiency; testing the Bell inequalities. Found. Phys. \textbf{19}, 453--478 (1989).

\bibitem{f94} Fine, A.: The Shaky Game: Einstein, Realism and the Quantum Theory. University of Chicago Press, Chicago, IL (1994).

\bibitem{a05} Accardi, L.: Some loopholes to save quantum nonlocality. In: Adenier, G., Khrennikov, A. (eds.) Foundations of Probability and Physics-3, pp. 1--20. American Institute of Physics, Ser. Conference Proceedings 750, Melville (2005).

\bibitem{k09} Khrennikov, A.: Interpretations of Probability. De Gruyter, Berlin (1998, 2009).

\bibitem{k05} Khrennikov, A., Smolyanov, O.G., Truman, A.: Kolmogorov probability spaces describing Accardi models for quantum correlations. Open. Syst. Inf. Dyn. {\bf 12}, 371--384 (2005).

\bibitem{h03} Hess, K., Philipp, W.: Exclusion of time in Mermin’s proof of Bell-type inequalities. In: Khrennikov, A. (ed.) Quantum Theory: Reconsideration of Foundations-2, pp. 243--254. V\"{a}xj\"{o} University Press, Ser. Math. Model. 10, V\"{a}xj\"{o} (2003).

\bibitem{h05} Hess, K., Philipp, W.: Bell's theorem: critique of proofs with and without inequalities. In: Adenier, G., Khrennikov, A. (eds.) Foundations of Probability and Physics-3, pp. 150--155. American Institute of Physics, Ser. Conference Proceedings 750, Melville (2005).

\bibitem{k12} Khrennikov, A: Quantum probabilities and violation of CHSH-inequality from classical random signals and threshold type detection scheme. Prog. Theor. Phys. {\bf 128}, 31--58 (2012).

\bibitem{k12a} Khrennikov, A.: Born’s rule from measurements of classical signals by threshold detectors which are properly calibrated. J. Mod. Opt. {\bf 59}, 667--678 (2012).

\bibitem{k12b} Khrennikov, A.: Born's rule from measurements of classical random signals under the assumption of ergodicity at the subquantum time scale. Op. Sys. Inf. Dyn. \textbf{19}, 48--65 (2012).

\bibitem{a09} Adenier, G.: Violation of Bell inequalities as a violation of fair sampling in threshold detectors. In: Accardi, L., {\it et al.} (eds.) Foundations of Probability and Physics-5. pp. 8--18. American Institute of Physics,  Ser. Conference Proceedings 1101, Melville (2009).

\bibitem{knew1} Khrennikov, A.: Towards new Grangier type experiments. Ann. Phys. {\bf 327}, 1786--1802 (2012).

\bibitem{knew2} Khrennikov, A.: Role of detectors and their proper calibration in inter-relation between classical and quantum optics. Opt. Eng. {\bf 51}(6), 069001 (2012).

\bibitem{kprep01} Khrennikov, A.: CHSH inequality: Quantum probabilities as classical conditional probabilities. {\it ArXiv: 1406.4886v1 [quant-ph]} (2014).

\bibitem{kprep02} Khrennikov, A.: Unconditional quantum correlations do not violate Bell's inequality. {\it ArXiv: 1503.08016v1 [quant-ph]} (2015).

\bibitem{kprep03} Khrennikov, A.: Classical probabilistic realization of ``Random Numbers Certified by Bell's Theorem''. {\it ArXiv: 1501.03581v1 [quant-ph]} (2015).



\end{thebibliography}
\end{document}